\documentclass[aps,prd,twocolumn,10pt,nofootinbib,superscriptaddress,preprintnumbers,floatfix]{revtex4-1}
\pdfoutput=1 
\usepackage[utf8]{inputenc} 
\usepackage{amsmath} 
\usepackage{amsfonts}
\usepackage{amsthm}
\usepackage{amssymb}
\usepackage{graphicx} 
\usepackage{color} 
\usepackage[normalem]{ulem} 
\usepackage{slashed} 
\usepackage{physics} 
\usepackage{mathrsfs} 
\usepackage{tikz}  
\usepackage[compat=1.1.0]{tikz-feynman} 
\usepackage{hyperref} 

\graphicspath{{./}{./figs/}}

\newcommand{\feynalign}[1]{
    \begin{gathered}
    \begin{tikzpicture}[scale=0.5, transform shape]
    \begin{feynman}
    #1
    \end{feynman}
    \end{tikzpicture}
    \end{gathered}
}
\newcommand{\tadpoletree}[2]{
    \vertex (a) at (0,0);
    \vertex[#2] (b) at (0,1) {};
    \diagram*{
    (a) -- [#1] (b) 
    };
}
\newcommand{\tadpole}[1]{
    \vertex (a) at (0,0);
    \vertex (b) at (0,0.5);
    \diagram*{
    (a) -- [#1] (b) 
    };
    \draw[#1] (b) arc [start angle=-90, end angle=270, radius=0.7cm];
}
\newcommand{\tadpoleinsertion}[3]{
    \vertex (a) at (0,0);
    \vertex (b) at (0,0.5);
    \vertex[#3] (c) at (0,1.9) {};
    \diagram*{
    (a) -- [#1] (b) -- [#2, half left] (c) -- [#2, half left] (b),
    };
}
\newcommand{\selfenergyfour}[2]{
    \vertex (a) at (0,0);
    \vertex (b) at (0.5,0.5);
    \vertex (c) at (1,0);
    \diagram*{
    (a) -- [#1] (b) -- [#1] (c)
    };
    \draw[#2] (b) arc [start angle=-90, end angle=270, radius=0.7cm];
}
\newcommand{\selfenergythree}[2]{
    \vertex (a) at (0,0);
    \vertex (b) at (0.5,0);
    \vertex (c) at (1.9,0);
    \vertex (d) at (2.4,0);
    \diagram*{
    (a) -- [#1] (b) -- [#2, half left] (c) -- [#2, half left] (b),
    (c) -- [#1] (d),
    };
}
\newcommand{\selfenergytree}[2]{
    \vertex (a) at (0,0);
    \vertex[#2] (b) at (0.7,0.7) {};
    \vertex (c) at (1.4,0);
    \diagram*{
    (a) -- [#1] (b) -- [#1] (c)
    };
}

\newcommand{\sunset}[2]{
    \vertex (a) at (0,0);
    \vertex (b) at (0.5,0);
    \vertex (c) at (1.9,0);
    \vertex (d) at (2.4,0);
    \diagram*{
    (a) -- [#1] (b) -- [#1, half left] (c) -- [#2, half left] (b),
    (c) -- [#1] (d), (b) -- [#2] (c),
    };
}

\newcommand{\sunsetpieceone}[1]{
    \tikzfeynmanset{every vertex={dot,minimum size=3mm}}
    \vertex[] (a) at (0,0) {};
    \vertex (b) at (1.2,0);
    \vertex[] (d) at (2.4,0) {};
    \diagram*{
    (a) -- [#1] (b),
    (b) -- [#1] (d),
    };
}

\newcommand{\sunsetpiecetwo}[1]{
    \tikzfeynmanset{every vertex={dot,minimum size=3mm}}
    \vertex[] (a) at (0,0) {};
    \vertex (b) at (1.2,0);
    \vertex[] (d) at (2.4,0) {};
    \diagram*{
    (a) -- [#1] (b),
    (b) -- [#1] (d),
    };
    \draw[#1] (b) arc [start angle=-90, end angle=270, radius=0.5cm];
}

\newcommand{\selfenergyoneloop}[3]{
    \vertex (a) at (0,0);
    \vertex (b) at (0.5,0.5);
    \vertex[#3] (d) at (0.5,0.5) {};
    \vertex (c) at (1,0);
    \diagram*{
    (a) -- [#1] (b) -- [#1] (c)
    };
    \draw[#2] (b) arc [start angle=-90, end angle=270, radius=0.7cm];
}
\newcommand{\selfenergyoneloopmassinsert}[3]{
    \vertex (a) at (0,0);
    \vertex (b) at (0.5,0.5);
    \vertex (c) at (1,0);
    \vertex[#3] (d) at (0.5,1.9) {};
    \diagram*{
    (a) -- [#1] (b) -- [#1] (c),
    (b) -- [#2, half left] (d) -- [#2, half left] (b)
    };
}

\newcommand{\selfenergysnowman}[1]{
    \vertex (a) at (0,0);
    \vertex (b) at (0.5,0.5);
    \vertex (c) at (1,0);
    \vertex (d) at (0.5,1.5);
    \diagram*{
    (a) -- (b) -- (c)
    };
    \draw[dashed] (b) arc [start angle=-90, end angle=270, radius=0.5cm];
    \draw[#1] (d) arc [start angle=-90, end angle=270, radius=0.5cm];
}

\newcommand{\selfenergysnowmanmod}[3]{
    \vertex (a) at (0,0);
    \vertex (b) at (0.5,0.5);
    \vertex (c) at (1,0);
    \vertex (d) at (0.5,1.5);
    \diagram*{
    (a) --[#1] (b) --[#1] (c)
    };
    \draw[#2] (b) arc [start angle=-90, end angle=270, radius=0.5cm];
    \draw[#3] (d) arc [start angle=-90, end angle=270, radius=0.5cm];
}

\newcommand{\sunsettwo}[2]{
    \vertex (a) at (0,0);
    \vertex (b) at (0.5,0);
    \vertex (c) at (1.9,0);
    \vertex (d) at (2.4,0);
    \diagram*{
    (a) -- [#1] (b) -- [#2, half left] (c) -- [#2, half left] (b),
    (c) -- [#1] (d), (b) -- [#1] (c),
    };
}

\newcommand{\fourpointtree}[1]{
    \vertex (a) at (0,-0.7);
    \vertex (b) at (0,0.7);
    \vertex (c) at (0.7,0);
    \vertex (d) at (1.4,-0.7);
    \vertex (e) at (1.4,0.7);
    \vertex[#1] (f) at (0.7,0) {};
    \diagram*{
    (a) -- (c),
    (b) -- (c), (c) -- (d), (c) -- (e),
    };
}

\newcommand{\fourpointtreemodded}[1]{
    \vertex (a) at (0,-0.7);
    \vertex (b) at (0,0.7);
    \vertex (c) at (0.7,0);
    \vertex (d) at (1.4,-0.7);
    \vertex (e) at (1.4,0.7);
    \vertex[#1] (f) at (0.7,0) {};
    \diagram*{
    (a) --[dashed] (c),
    (b) --[dashed] (c), (c) -- (d), (c) --(e),
    };
}

\newcommand{\fourpointcorrection}{
    \vertex (a) at (0,-0.5);
    \vertex (b) at (0,0.5);
    \vertex (c) at (0.5,0);
    \vertex (d) at (1.9,0);
    \vertex (e) at (2.4,-0.5);
    \vertex (f) at (2.4,0.5);
    \diagram*{
    (a) -- (c) -- [dashed, half left] (d) -- [dashed, half left] (c),
    (b) -- (c), (d) -- (e), (d) -- (f),
    };
}

\newcommand{\fourpointcorrectionmod}[1]{
    \vertex (a) at (0,-0.5);
    \vertex (b) at (0,0.5);
    \vertex (c) at (0.5,0);
    \vertex (d) at (1.9,0);
    \vertex (e) at (2.4,-0.5);
    \vertex (f) at (2.4,0.5);
    \diagram*{
    (a) --[dashed] (c) -- [#1, half left] (d) -- [#1, half left] (c),
    (b) --[dashed] (c), (d) -- (e), (d) -- (f),
    };
}

\newcommand{\oneloopbubble}{
    \vertex (a) at (0,0);
    \draw[dashed] (a) arc [start angle=-90, end angle=270, radius=0.7cm];
}

\newcommand{\twoloopbubble}{
    \vertex (a) at (0.5,0.5);
    \vertex (b) at (0.5,1.5);
    \draw[dashed] (a) arc [start angle=-90, end angle=270, radius=0.5cm];
    \draw[dashed] (b) arc [start angle=-90, end angle=270, radius=0.5cm];
}

\newcommand{\sunsetbubble}{
    \vertex (b) at (0.5,0);
    \vertex (c) at (1.9,0);
    \diagram*{
    (b) -- [dashed, half left] (c) -- [dashed, half left] (b),
    , (b) -- (c),
    };
}

\newcommand{\cammassinsert}{
    \vertex (b) at (0.5,0.5);
    \vertex[crossed dot] (d) at (0.5,1.9) {};
    \diagram*{
    (b) -- [dashed, half left] (d) -- [dashed, half left] (b)
    };
}

\newcommand{\shapechangeScaleShifter}{
    \vertex (a) at (0.0,0.0);
    \vertex (b) at (1.0,0.0);
    \vertex[crossed dot] (c) at (-1.4,0.0) {};
    \vertex[crossed dot] (d) at (2.4,0.0) {};
    \diagram*{
    (a) -- [dashed, half left] (c) -- [dashed, half left] (a);
    (b) -- [dashed, half left] (d) -- [dashed, half left] (b);
    (a) -- (b)
    };
}

\newcommand{\shapechangenucl}{
    \vertex (a) at (0.0,0.0);
    \vertex (b) at (1.0,0.0);
    \draw[dotted, thick] (a) arc [start angle=0, end angle=360, radius=0.7cm];
    \draw[dotted, thick] (b) arc [start angle=-180, end angle=180, radius=0.7cm];
    \diagram*{
    (a) -- (b)
    };
}

\newcommand{\shapechangesaddle}{
    \vertex (a) at (0.0,0.0);
    \vertex (b) at (0.7,0.0);
    \draw[dotted, thick] (b) arc [start angle=-180, end angle=180, radius=0.7cm];
    \diagram*{
    (a) -- (b)
    };
}

\newcommand{\oneloopbubbledotted}{
    \vertex (a) at (0,0);
    \draw[dotted, thick] (a) arc [start angle=-90, end angle=270, radius=0.7cm];
}

\newcommand{\blob}{
    \fill[gray!50] (0,0) circle (0.5cm);
    \draw (0,0) circle (0.5cm)      node[right=.7cm, above=.4cm] {};
}

\newcommand{\blobshape}{
    \fill[gray!50] (-0.5,0) circle (0.5cm);
    \draw (-0.5,0) circle (0.5cm);
    \fill[gray!50] (1.5,0) circle (0.5cm);
    \draw (1.5,0) circle (0.5cm);
    \vertex (a) at (0.0,0.0);
    \vertex (b) at (1.0,0.0);
    \diagram*{
    (a) -- (b)
    };
}

\newcommand{\mr}[1]{\mathrm{#1}}
\newcommand{\mb}[1]{\mathbf{#1}}
\newcommand{\mbb}[1]{\mathbb{#1}}
\newcommand{\mc}[1]{\mathcal{#1}}
\newcommand{\ms}[1]{\mathscr{#1}}
\newcommand{\rmi}[1]{{\mbox{\scriptsize #1}}}
\newcommand{\rmii}[1]{{\mbox{\tiny\rm{#1}}}}

\newcommand{\cb}{\phi_{\rmii{CB}}}
\newcommand{\sn}{S_{\rmi{nucl}}}
\newcommand{\approachtitle}[1]{\vspace{0.1cm} \noindent {\bf \emph{#1} --}}

\newcommand{\F}{ F_{\rmi{eff}} }
\newcommand{\kap}{ \kappa }
\newcommand{\LamTh}{ \Lambda_{\rm therm} }
\newcommand{\LamNuc}{ \Lambda_{\rm nucl} }
\newcommand{\mNuc}{ m_{\rm nucl} }
\newcommand\sumint[1]{\hbox{$\sum$}\!\!\!\!\!\!\!\int_{#1}}
\newcommand{\cou}[1]{\delta_{#1}}
\newcommand\MSbar{$\overline{\text{MS}}$}
\newcommand\gE{\gamma_{\text{E}}}
\newcommand{\Chi}{\mathrm{X}}
\newcommand{\blam}{\Lambda}
\newcommand{\bmu}{\mu}

\newcommand{\rNonlocal}{ \tilde{r} } 
\newcommand{\phiNonlocal}{ \phi_{\rmi{nonlocal}} }

\newcommand{\Helsinki}{\affiliation{
    Department of Physics and Helsinki Institute of Physics,
    P.O.\ Box 64,
    FI-00014 University of Helsinki,
    Finland
}}
\newcommand{\Nottingham}{\affiliation{
    School of Physics and Astronomy,
    University Park,
    University of Nottingham,
    Nottingham NG7 2RD,
    United Kingdom
}}

\begin{document}

\title{Effective field theory approach to thermal bubble nucleation}
\date{August 9, 2021} 

\author{Oliver Gould}
\email{oliver.gould@nottingham.ac.uk}
\Nottingham
\Helsinki

\author{Joonas Hirvonen}
\email{joonas.o.hirvonen@helsinki.fi}
\Helsinki

\begin{abstract}
The standard vacuum bounce formalism suffers from inconsistencies when applied to thermal bubble nucleation, for which ad hoc workarounds are commonly adopted.
Identifying the length scales on which nucleation takes place, we demonstrate how the construction of an effective description for these scales naturally resolves the problems of the standard vacuum bounce formalism.
Further, by utilising high-temperature dimensional reduction, we make a connection to classical nucleation theory.
This offers a clear physical picture of thermal bubble nucleation, as well as a computational framework which can then be pushed to higher accuracy.
We demonstrate the method for three qualitatively different quantum field theories.
\end{abstract}

\preprint{HIP-2020-19/TH}
\maketitle

\tableofcontents


\section{Introduction} \label{sec:introduction}

Since the hot Big Bang, the universe may have passed through a number of different phases.
The dramatic possibility of a strong first-order phase transition arises in many theories beyond the Standard Model.
In particular such a transition is a necessary ingredient for successful electroweak baryogenesis~\cite{Kuzmin:1985mm,Shaposhnikov:1986jp,Shaposhnikov:1987tw,Morrissey:2012db}.

Since the discovery of gravitational waves by LIGO~\cite{Abbott:2016blz},
the subject of cosmological first-order phase transitions has gathered fresh interest due to the stochastic gravitational wave background which they would produce
(for recent reviews see Refs.~\cite{Caprini:2018mtu,Caprini:2019egz,Hindmarsh:2020hop}).
The stochastic gravitational wave background may be observable by current and near-future gravitational-wave detectors~\cite{Arzoumanian:2020vkk,Audley:2017drz,Kawamura:2011zz,Harry:2006fi,Guo:2018npi}.
This offers a new window into the early universe,
including of dark sectors uncoupled to the Standard Model~\cite{Schwaller:2015tja,Croon:2018erz,Breitbach:2018ddu}.

First-order phase transitions proceed through the nucleation of bubbles of a stable phase, which grow until they eventually supplant a pre-existing metastable phase.
As a consequence, the study of bubble nucleation lies at the heart of the theory of first-order phase transitions.
The rate of bubble nucleation determines many important physical properties of first-order phase transitions.
In particular, the peak frequency and amplitude of the gravitational wave spectrum depend directly on the rate of bubble nucleation.

From an observation of a stochastic gravitational wave background and its spectrum it is possible to learn about the process which produced it.
In principle one could extract quantitative information about the underlying particle physics models.
For this, accurate and reliable predictions of the bubble nucleation rate and related phase transition parameters are essential.
Yet typically, calculations of these parameters are subject to huge theoretical uncertainties:
a multiplicative uncertainty in the gravitational wave peak amplitude of many orders of magnitude~\cite{Croon:2020cgk,Gould:2021oba}.
By pushing to higher perturbative orders, it is possible to dramatically reduce these uncertainties for purely equilibrium quantities, such as the critical temperature and latent heat~\cite{Kainulainen:2019kyp,Niemi:2020hto,Niemi:2021qvp}.
However, for the bubble nucleation rate this possibility is still out of reach, and as a consequence it is a limiting source of uncertainty in the gravitational wave spectrum.
Thus, a better understanding of how to calculate the rate of thermal bubble nucleation is critical.

The modern theory of bubble nucleation was initiated by Langer in the late 1960s~\cite{Langer:1967ax,Langer:1969bc,langer1974metastable}, in the context of classical statistical mechanics.
This built upon earlier work by Kramers~\cite{Kramers:1940zz} regarding the escape rate of a particle trapped in a potential well (see also related work by Zel'dovich~\cite{zeldovich1942theory,lifshitz1981kinetics}).
An important development was made by Coleman in 1977~\cite{Coleman:1977py},
who derived an analogous formalism for vacuum decay in relativistic quantum field theory (QFT) at zero temperature.
Shortly afterwards, Linde outlined how vacuum decay would generalise to nonzero temperature~\cite{Linde:1980tt,Linde:1981zj}.

However, a lack of clarity on the details of this generalisation to nonzero temperature has meant that concrete calculations are plagued by internal inconsistencies,
as has been observed by many authors~\cite{langer1974metastable,Buchmuller:1992rs,Buchmuller:1993bq,Bodeker:1993kj,Gleiser:1993hf,Alford:1993br,Berges:1996ib,Berges:1996ja,Litim:1996nw,Surig:1997ne,Strumia:1998nf,Strumia:1998qq,Strumia:1998vd,Munster:2000kk,Strumia:1999fq,Garbrecht:2015yza,Ai:2018rnh,Croon:2020cgk} (see also Refs.~\cite{Weinberg:1992ds,Garbrecht:2015yza} for related issues at zero temperature).
In short, there is an apparent catch-22 because one must integrate out thermal fluctuations in order to solve for the bubble configuration, and yet one must already know the bubble configuration in order to integrate over the fluctuations about it.
Naive attempts to resolve this apparent catch-22 lead to erroneously double-counting degrees of freedom (DoFs), stray imaginary parts and an uncontrolled derivative expansion; see Sec.~3.5 of Ref.~\cite{Croon:2020cgk} for a fuller explanation.
We note that the same problem arises in a variety of other contexts, typically where the effective action, rather than the effective potential, is involved.

An important clue to solving these problems is the classical nature of nucleation for high-temperature phase transitions which are dictated by long-wavelength physics.
In a study of the quantum mechanical escape rate, it was shown by Affleck~\cite{Affleck:1980ac} that, as the temperature rises, the process changes (relatively abruptly) from quantum tunnelling to thermal over-barrier escape, with the latter given by a classical thermal calculation.
This behaviour is also expected to occur for the bubble nucleation rate in QFT, a suggestion reinforced by the known classicalisation of long-wavelength modes at high temperatures (see for example~\cite{Mueller:2002gd,Greiner:1996dx,Aarts:1996qi,Aarts:1997kp,Bodeker:1996wb}).
Therefore, when it is the long-wavelength modes which undergo nucleation, the quantum tunnelling picture should be supplanted by an effective classical picture of bubble nucleation.

Nonetheless, the precise link has remained obscure between Langer's classical formalism and the study of thermal bubble nucleation within QFT.
We aim to clarify this link here, demonstrating it explicitly.
However, we will leave one notable missing piece, the calculation of the dynamical prefactor (discussed in Sec.~\ref{sec:background}), for future work.
Our work utilises insights from E. Weinberg's work on radiatively-induced vacuum transitions~\cite{Weinberg:1992ds},
as well as the literature on the application of effective field theory to the thermodynamics of QFTs, high temperature dimensional reduction~\cite{Farakos:1994kx,Kajantie:1995dw,Braaten:1995cm,Braaten:1995jr}.
There are also connections between our work and Ref.~\cite{Berera:2019uyp}, which recast Langer's formalism for a scalar field Langevin equation.
Finally, we would like to highlight Refs.~\cite{Moore:2000jw,Moore:2001vf} in which a lattice formulation of bubble nucleation was presented, based on similar underlying ideas.

Effective field theory (EFT) plays a central role in our thesis, for two apparently distinct reasons.
First, EFT resolves the aforementioned catch-22.
To do this, there must be a split between the shorter wavelength modes which contribute to the background for the critical bubble, and the longer wavelength fluctuations of the critical bubble itself.
Further, there must be a hierarchy of scales between these two sets of modes if a local description of bubble nucleation is to exist, in which case EFT is the natural tool to make the split.
Thus EFT provides a solution to the problems and inconsistencies present in existing calculations of the thermal bubble nucleation rate.

Secondly, high-temperature EFTs~\cite{Farakos:1994kx,Kajantie:1995dw,Braaten:1995cm,Braaten:1995jr} provide a bridge to classical nucleation theory.
In fact, an effective description is assumed as the starting point of classical nucleation theory~\cite{Langer:1969bc} (as explained in Ref.~\cite{langer1974metastable}), hence EFT provides the possibility to derive this starting point from an underlying QFT.
We inspect the classical nucleation theory and note that the exponentially large contributions are contained in its statistical part, calculable with a classical equilibrium description.
The same structure can be identified within the imaginary time formalism of thermal QFT, by constructing an EFT for the nucleating degrees of freedom.
In sum, high-temperature EFTs provide the correct classical effective description for thermal bubble nucleation.

In addition to self-consistency (or perhaps following in consequence), the approach presented in this paper yields important benefits.
First, dependence on the renormalisation scale is minimal, as the dependence cancels order-by-order.
This is noteworthy because typical computations in the literature generally show a strong unphysical scale dependence~\cite{Croon:2020cgk,Gould:2021oba}.
Second, although we do not explicitly study a model involving gauge fields here, the approach will in general yield order-by-order gauge invariant results, as both dimensional reduction and the calculation within the three-dimensional EFT are gauge invariant~\cite{Croon:2020cgk}.
The latter follows because the calculation within the EFT is equivalent to that of the vacuum decay rate for a QFT in three spacetime dimensions, and hence its gauge-invariance is guaranteed by the Nielsen-Fukuda-Kugo identities~\cite{Nielsen:1975fs,Fukuda:1975di}; see for example Refs.~\cite{Metaxas:1995ab,Metaxas:2000cw,Plascencia:2015pga,Arunasalam:2021zrs}.

In Sec.~\ref{sec:background} we review the general theory of thermal bubble nucleation, showing how the bubble nucleation rate splits into dynamical and statistical parts.
Following Langer, we start from an effective description of the nucleating degrees of freedom.
In Sec.~\ref{sec:formalism} we outline a general formalism for the derivation of this effective description from an underlying QFT.
In this we work from first principles, drawing from the literature on effective field theory.
The formalism is then applied in Sec.~\ref{sec:one_scalar} to a simple Yukawa theory in which only one scale separates the microscopic QFT from the effective description of the nucleating degrees of freedom.
In Sec.~\ref{sec:two_scalar} we demonstrate the formalism in a more complicated two-scalar theory, in which there are two scale hierarchies.
A third example follows in Sec.~\ref{sec:scaleshifters}, the cubic anisotropy model, which demonstrates
an important complication to the EFT method which is present in, for example, gauge-Higgs theories.
Finally, in Sec.~\ref{sec:discussion}, we compare our formalism to others in the literature, and suggest directions for future progress.


\section{Classical nucleation theory} \label{sec:background}

In this section, we review classical nucleation theory, as put forward by Langer~\cite{Langer:1967ax,Langer:1969bc,langer1974metastable}.
In addition, we will draw attention to a key result: the factorisation of the nucleation rate into dynamical and statistical parts.
Classical nucleation theory can be justified from an underlying QFT when the low energy modes are the ones undergoing nucleation, which is natural in a weakly coupled theory.
Bosonic modes with energies much lower than the temperature $E\ll T$ are Bose enhanced and become highly occupied $n_B=1/(e^{E/T}-1)\approx T/E \gg 1$, leading to their classicalisation~\cite{Mueller:2002gd,Greiner:1996dx,Aarts:1996qi,Aarts:1997kp,Bodeker:1996wb}.
Unequal time correlation functions of these modes can be computed equally well in either a classical or a quantum theory, up to some order in the couplings.
This justifies the use of classical nucleation theory in the saddle-point approximation (at one-loop order within the classical EFT).
Quantum corrections to the nucleation rate should also be expected to be calculable classically, by matching to the quantum theory (see for example~\cite{Bodeker:1996wb}).

Those readers who are familiar with classical nucleation theory, or who are willing to take on faith the split of the nucleation rate into a product of statistical and dynamical parts, may wish to skip ahead to Sec.~\ref{sec:formalism}.

A metastable phase decays into a stable phase via the nucleation and growth of bubbles.
As the phase is only metastable, some thermal fluctuations are large enough to escape into the stable phase and begin to grow.
These large enough fluctuations, which interpolate between the two phases, are the bubbles.

We will focus on the calculation of the rate of bubble nucleation.
It is well defined only if the timescale for nucleation is much longer than the timescales of particle scatterings in the metastable phase.
This is necessary for the system to be able to thermalise in the metastable phase.
As we will see below, this hierarchy of timescales can follow naturally from the Boltzmann suppression of bubble configurations.

Following Langer~\cite{Langer:1967ax,Langer:1969bc,langer1974metastable}, we will begin by assuming that there exists some effective, coarse-grained description of the nucleating system.
In particular, we will assume that it can be described via an effective classical field, $\phi(\mb{x})$, and an effective free energy, $\F[\phi]$, which describes its dynamics.
Here, in principle, the effective field may have any number of degrees of freedom at each spatial point, $\mb{x}$, though we will suppress any corresponding internal indices.
The effective free energy is given schematically by~\cite{langer1974metastable}
\begin{equation}\label{eq:LangerFree}
    e^{-\beta \F[\phi]}\; = \sum_{\left[\substack{\mr{constrained}\\ \mr{microscopic\ variables}}\right]} e^{-\beta H} \,,
\end{equation}
where $H$ is the Hamiltonian describing the system at the shortest scales.%
\footnote{
Note that Eq.~\eqref{eq:LangerFree} assumes that there are no relevant chemical potentials.
Otherwise, they would enter the formula.
} The main focus of this paper, and the task for Sec.~\ref{sec:formalism}, is to 
outline how to carry out this schematic sum starting from an underlying description in terms of a relativistic QFT.

The effective description should describe the dynamics of the system on the length and time scales relevant for bubble nucleation.
Admissible choices are such that the effective free energy exhibits the correct phase structure, with two phases separated by a potential barrier, and that the effective field can describe bubble configurations.
The former condition states that the fluctuations inducing the transition have been integrated out, and the latter states that the bubbles themselves have not been coarse-grained out of the description.

In Langer's analysis~\cite{Langer:1969bc} the dynamical equations for $\phi(\mb{x})$ were assumed to be stochastic Hamiltonian equations, and in Ref.~\cite{Berera:2019uyp} an equivalent description in terms of Langevin equations was adopted.
A Langevin-like description is in fact formally equivalent to the Schwinger-Dyson equations~\cite{Gautier:2012vh} for the time evolution of correlation functions with a good classical limit (i.e.\ for expectation values of anti-commutators of operators), though generically such descriptions are nonlocal in both space and time.
Markovian (and hence local in time) Langevin equations are nevertheless applicable to the slowest evolving DoFs, for which the faster DoFs act as a thermal bath, and a source of stochasticity and damping~\cite{Greiner:1996dx,Bodeker:1999ey} (see also Refs.~\cite{Morikawa:1986rp,Gleiser:1993ea,Berera:1998gx,Yokoyama:2004pf,Berera:2007qm}).
For the longest wavelength fluctuations, these Langevin equations may also be local in space~\cite{Bodeker:1999ey}.
In principle one could extend classical nucleation theory to alternative dynamical descriptions, such as the hard thermal loops effective description for soft gauge fields~\cite{Blaizot:1993zk,Blaizot:1993be}.
However, in the following we will assume the existence of generalised stochastic Hamiltonian equations of the form given in Ref.~\cite{Langer:1969bc} for the dynamics of the field $\phi(\mb{x})$.

Classicality entails a probabilistic or statistical description of the system.
The calculation of the bubble nucleation rate can then be formulated as a statistical initial value problem, in which the system starts in the metastable region, and one is interested in the rate at which the probability distribution leaks over into the stable region.

\begin{figure}[t]
  \centering
  \includegraphics[width=0.95\linewidth]{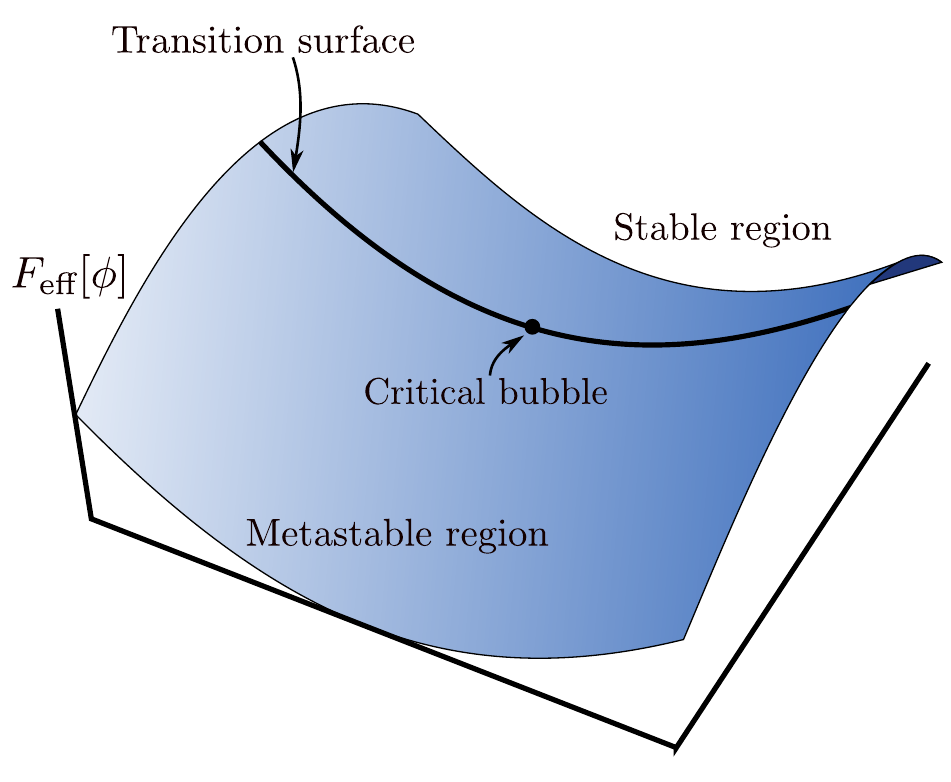}  
  \caption{
  Diagram showing a two-dimensional slice of the configuration space of a nucleating system in the vicinity of the saddle point.
  The dot on the transition surface corresponds to the critical bubble, which is the lowest point on the transition surface.
  }
  \label{fig:critical_bubble}
\end{figure}

From the effective free energy one can define the {\em transition surface} in the space of field configurations.
It can be defined as the separatrix between gradient flows on $\F[\phi]$ which lead to the two different phases, marking the high point between them, illustrated in Fig.~\ref{fig:critical_bubble}.
The transition surface bifurcates the space of field configurations, separating the two phases.

Physically, crossing the transition surface means nucleating a bubble.
This must happen locally in space, as the exponential suppression of the Boltzmann factor grows linearly with volume.
In fact, from the assumption that bubble nucleation is slow, one need only consider the nucleation of a single bubble, as the probability of producing two bubbles near each other is additionally exponentially suppressed.
This can be formalised with a virial expansion (see for example Ref.~\cite{pathria32statistical}), in this context sometimes called a dilute instanton approximation~\cite{Callan:1977gz}.

The critical bubble, denoted with $\cb$, is the gatekeeper to the stable phase, being the least suppressed effective field configuration on the transition surface, shown in Fig.~\ref{fig:critical_bubble}.
Consequently, it is also a stationary point of the effective free energy,
\begin{equation}\label{eq:StationaryPoint}
    \fdv{\F}{\phi}\eval_{\phi=\cb}=0\,.
\end{equation}
The boundary conditions are such that the field is in the metastable state at spatial infinity.
Arbitrary effective field configurations on the transition surface are typically much more exponentially suppressed.
Therefore, one expects that all nucleating bubbles can be handled as slightly deformed critical bubbles.

This idea can be formulated as a saddle-point approximation over the transition surface, in which the effective free energy is approximated to second order in deviations about the critical bubble:
\begin{align}
    \F[\cb+\varphi]&\approx \F[\cb]\nonumber\\
    &+\frac{1}{2}\int_{\mb{x}}\int_{\mb{y}}\varphi(\mathbf{x}) \F''[\cb](\mathbf{x},\mathbf{y}) \varphi(\mathbf{y}) ,\label{eq:SecondOrder}\\
    \F''[\cb](\mathbf{x},\mathbf{y}) =& \frac{\delta^2\F}{\delta\phi(\mathbf{x})\delta\phi(\mathbf{y})}\eval_{\phi=\cb}\,,\label{eq:FluctuationOperator}
\end{align}
where we have introduced $\int_{\mb{x}}\equiv\int\dd^3x$ for integration over space, and the linear term is zero due to Eq.~\eqref{eq:StationaryPoint}.

The fluctuation operator, $\F''[\cb]$ from Eq.~\eqref{eq:FluctuationOperator}, plays an important role in the saddle-point approximation, but also in understanding the transition surface and the direction of growth for a bubble within configuration space.
Crucially, the operator has one and only one negative eigenvalue.%
\footnote{
The dimensionality of the transition surface is one less than the configuration space. The critical bubble is the lowest point on the surface, therefore only one negative eigenmode orthogonal to the surface is admitted~\cite{Coleman:1987rm}.
}
The direction in configuration space corresponding to the negative eigenmode gives the direction of decreasing effective free energy, and hence the direction of bubble growth.
This direction is orthogonal to the transition surface, shown in Fig.~\ref{fig:critical_bubble}.

In addition to the negative eigenvalue $\F''[\cb]$ has one zero eigenvalue for each symmetry of $\F[\phi]$ broken by the critical bubble.
Translational symmetry is necessarily broken by the critical bubble, resulting in the rate being proportional to the volume of the system.
Broken internal symmetries also lead to zero eigenvalues.
The remaining eigenvalues are positive.

Mathematically, Langer defined the rate of bubble nucleation, $\Gamma$, as the flux of probability over the transition surface~\cite{Langer:1969bc},
\begin{align}
    \Gamma&=\int_{\mr{TS}}\, J \cdot dS_\perp\,,\label{eq:rateDefinition}
\end{align}
following earlier work by Kramers~\cite{Kramers:1940zz} on the {\em escape problem}.%
\footnote{There is freedom in choosing the integration surface in the full phase space, but this definition utilising the transition surface is the most convenient for the split into statistical and dynamical parts which follows.}
Here $J$ denotes the probability current, a vector field in the phase space of the effective description, and $dS_\perp$ denotes the surface element perpendicular to the transition surface, $\mr{TS}$.
The probability current can be defined by the continuity equation for the probability density.

This definition differs from $-2\Im F$, which has also been suggested as a definition for the thermal nucleation rate~\cite{Linde:1980tt,Linde:1981zj}, by analogy with the rate of vacuum decay at zero temperature $\Gamma_{\rm vac} = -2\Im E_{\rm vac}$.
However, while $\Im E_{\rm vac}$ can be directly linked to the process of time evolution in the false vacuum state~\cite{Schwinger:1951nm}, this is not the case for $\Im F$ and time evolution near thermal equilibrium, for which the necessary analytic continuations are more subtle~\cite{Chou:1984es}.
On the other hand, within an effective classical description, the interpretation of Eq.~\eqref{eq:rateDefinition} is unambiguous.
Ref.~\cite{Affleck:1980ac} found that the quantum mechanical escape rate of a particle is given by $-2\Im F$ at low temperatures, and by Eq.~\eqref{eq:rateDefinition} at high temperatures, where an effective classical description indeed applies.

To determine the probability distribution and current, one must solve the appropriate coarse-grained dynamical equations.
In the saddle-point approximation, the problem can be formulated in the vicinity of the critical bubble, where Eq.~\eqref{eq:SecondOrder} holds.
The relevant boundary conditions are such that, far from the transition surface, the metastable state is in thermal equilibrium while the stable state is unpopulated.
This problem was solved by Langer~\cite{Langer:1969bc}, building on earlier work by Kramers~\cite{Kramers:1940zz}.
The solution describes a small constant flow of probability from the metastable phase to the stable phase, with the boundary conditions acting as source and sink.

Crucially, the solution for the phase space probability distribution, $\rho$, splits into a product of the equilibrium distribution and a nonequilibrium factor which depends only on a single direction in phase space,
\begin{equation}\label{eq:offEqDist}
    \rho = \sigma(u) \frac{e^{-\beta \F}}{Z_\rmi{meta}}\,,
\end{equation}
where $\sigma$ is the deviation from the thermal distribution.
The parameter $u$ is a linear combination of directions in phase space:
\begin{equation}\label{eq:linComb}
    u=(U,\,\Bar{U})\cdot
    	\begin{pmatrix}
    	\phi - \cb \\ \pi
    	\end{pmatrix}.
\end{equation}
Here we are using the notation of Ref.~\cite{Berera:2019uyp}, in which $\phi$ describes the configuration space of the effective theory, $\pi$ denotes the momentum conjugate to $\phi$ and $(U,\,\Bar{U})$ denotes a vector of coefficients in the space of linear fluctuations about $(\phi,\pi)=(\cb,0)$.
The vector $(U,\,\Bar{U})$ gives the single direction, in which the probability distribution is out of equilibrium. It also carries information on the direction of the net probability flow in the phase space.

The quadratic approximation to the effective free energy, Eq.~\eqref{eq:SecondOrder}, is symmetric under the following transformation
\begin{equation}\label{eq:mirrorsymmetry}
    \begin{pmatrix}
        \phi - \cb \\ \pi
    \end{pmatrix}
    \to
    -
    \begin{pmatrix}
        \phi - \cb \\ \pi
    \end{pmatrix}.
\end{equation}
This symmetry is also satisfied by the corresponding equations of motion, including noise and damping (\textit{cf.}\ e.g.\ Eq.~(3.1) in Ref.~\cite{Berera:2019uyp}).%
\footnote{
    A multiplicative noise term, such as appears in Refs.~\cite{Greiner:1996dx,Gleiser:1993ea,Berera:1998gx,Berera:2007qm}, would break this symmetry.
    However, this occurs beyond the order of the discussion as at leading order such a term is given by $\phi\xi\approx\cb\xi$.
}

The symmetry is broken by the boundary conditions, which ensure that
the probability is flowing from the metastable state to the stable state.
In the quadratic approximation, the only distinction between the stable and the metastable phases are the boundary conditions: $\sigma\to 1$ on the metastable side and $\sigma\to 0$ on the stable side.
The negative eigenmode determines the direction to the phases in the quadratic approximation.
Hence, the symmetry is broken in a direction which is a linear combination of the negative eigenmode and its conjugate momentum, whose subspace we will refer to as the negative eigensubspace.
Due to the symmetry, the vector $(U,\,\Bar{U})$ can only lie in the negative eigensubspace.

\begin{figure}[t]
  \centering
  \includegraphics[width=0.95\linewidth]{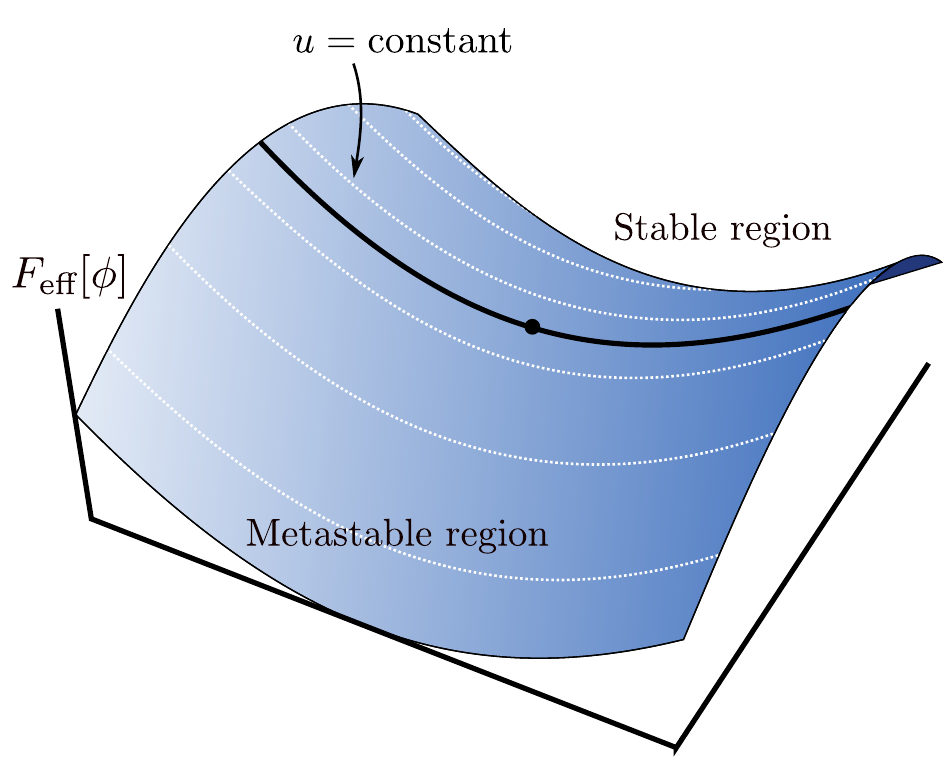}  
  \caption{
  Diagram building upon Fig.~\ref{fig:critical_bubble}, showing additionally lines of constant $u$, along which the stationary probability distribution is in thermal equilibrium; see Eqs.~\eqref{eq:offEqDist} and \eqref{eq:linComb}.
  }
  \label{fig:constLines}
\end{figure}

As the probability distribution is out of equilibrium only in the negative eigensubspace, it behaves thermally
\begin{equation}
    \rho\propto e^{-\beta \F}
\end{equation}
within the transition surface, as well as within other surfaces perpendicular to the negative eigensubspace, shown as white lines in Fig.~\ref{fig:constLines}.
Further, as there is no net probability flux in thermal equilibrium, the probability current must also lie in the negative eigensubspace.
It splits into an analogous product of terms:
\begin{equation}\label{eq:flow}
    J=\sigma_J(u)\, \frac{e^{-\beta \F}}{Z_\rmi{meta}} \,,
\end{equation}
where $\sigma_J(u)$ is a vector with constant direction in the negative eigensubspace of the phase space.%
\footnote{
    It is instructive here to compare Eq.~\eqref{eq:flow} with the approach of Ref.~\cite{Affleck:1980ac} (see also Ref.~\cite{Hirvonen:2020jud} for the field theory case).
    In that reference, the author did not solve any underlying dynamical equation for $\sigma_J(u)$.
    Instead an ansatz for the perpendicular component was proposed, amounting on the transition surface to $\sigma_{J,\phi_-}(u)\eval_{\mathrm{TS}}=\theta(\pi_-)\,\pi_-$, (\textit{cf.} below Eq.~\eqref{eq:nuclRate} for notation).
    This ansatz agrees with the solution of Kramers and Langer only in the limit of all dissipation constants being zero (\textit{cf.} Ref.~\cite{Hirvonen:2020jud}).
}

The nucleation rate, Eq.~\eqref{eq:rateDefinition}, is given by the integral of the probability flux over the transition surface, which can be written as
\begin{align}
    \Gamma&=\int\mathcal{D}\pi\mathcal{D}\phi\, \delta(\phi_-) J_{\phi_-} \,.\label{eq:nuclRate}
\end{align}
The variable $\phi_-$ is the coefficient of the negative eigenmode in the expansion $\phi=\sum_n \phi_n f_n$, where the functions $f_n$ are the normalised eigenmodes of $\F''[\cb]$.
Thus, $\phi_-$ is zero on the transition surface in the saddle-point approximation and the delta function in Eq.~\eqref{eq:nuclRate} enforces the integration over the transition surface.
The notation $J_{\phi_-}$ refers to the component of the probability current vector in the direction of the negative eigenmode, perpendicular to the transition surface.

From Eqs.~\eqref{eq:linComb}, \eqref{eq:flow} and \eqref{eq:nuclRate}, we can see that the only non-equilibrium integral is over the momentum conjugate to the negative eigenmode.
Hence, the nucleation rate formula splits into a product of statistical (equilibrium) and dynamical (nonequilibrium) parts,
\begin{align} \label{eq:divisionDynamicalStatistical}
    \Gamma &= \frac{\kappa}{2\pi} \; \Sigma \,,\\
    \Sigma &= \frac{\mathcal{N}}{Z_{\rmi{meta}}}\int \mathcal{D}\phi\,\delta(\phi_-) e^{-\beta \F[\phi]} \,. \label{eq:GeneralArea}
\end{align}
Here, the statistical part, denoted by $\Sigma$, is the Boltzmann weighted transition surface area normalised relative to the metastable phase, a purely equilibrium quantity.
To match the standard convention in the literature (e.g.\ Ref.~\cite{langer1973hydrodynamic}), we have multiplied it by a normalisation factor, $\mathcal{N}=(\beta|\lambda_-|/2\pi)^{-1/2}$, where $\lambda_-$ is the negative eigenvalue of the fluctuation operator in Eq.~\eqref{eq:FluctuationOperator}.
Also, the dynamical part, denoted by $\kappa$, now has a definite physical interpretation as the exponential growth rate of the critical bubble~\cite{Langer:1969bc}.
The split in Eq.~\eqref{eq:divisionDynamicalStatistical} is the main result of this section, and finding $\Sigma$ will be the focus of Sec.~\ref{sec:formalism}. 

The exponential suppression of the nucleation rate comes from the Boltzmann suppression of the configurations on the transition surface, whereas the more complicated non-equilibrium physics affects only the dimensionful prefactor, $\kappa$.
The result of integration over the conjugate momentum of the negative eigenmode is hidden into the dynamical part, $\kap$, in Eq.~\eqref{eq:divisionDynamicalStatistical} along with other information in $\sigma_J(u)$.
All the purely equilibrium behaviour is contained in $\Sigma$.

Actually, there is a slight mismatch between the free energies, $\F$, and partition functions, $Z_{\rmi{meta}}$, entering Eqs.~\eqref{eq:nuclRate} and \eqref{eq:GeneralArea}.
In the former equation, we have not yet performed the conjugate momentum integrals and the free energy contains quadratic dependence on them; see for example Refs.~\cite{Aarts:1996qi,Aarts:1997kp}.
In the latter, they have been integrated over, and no conjugate momentum dependence is left.
As discussed above, the contribution from the non-equilibrium integral over the momentum conjugate to the negative eigenmode is included in the dynamical part, $\kap$.
The rest of the conjugate momentum integrals, unaffected by the critical bubble background, cancel against the corresponding ones in $Z_{\rmi{meta}}$.
The free energy obtained in Sec.~\ref{sec:formalism}, via the imaginary time formalism, is akin to Eq.~\eqref{eq:GeneralArea}, containing no conjugate momenta, only field dependence.

Utilising the analysis in Ref.~\cite{Berera:2019uyp}, we can find the exponential growth rate of the bubbles in terms of the friction coefficient, $\eta$,
\begin{equation}
    \kappa=\sqrt{\abs{\lambda_-}+\eta^2/4}-\eta/2 \,,
\end{equation}
when the equation of motion is the standard Langevin equation for a real scalar field.
Here, we have continued to use the notation of Ref.~\cite{Langer:1969bc} for the negative eigenvalue, which differs from the convention in Ref.~\cite{Berera:2019uyp}.

Performing the saddle-point approximation for the statistical part, one finds
\begin{align}
    \Sigma &=\mc{V}\sqrt{\abs{\frac{\det (\beta \F''[\phi_\rmi{meta}]/2\pi)}{\det^\prime (\beta \F''[\cb]/2\pi)}}}\ e^{-\beta \Delta\F} \,.\label{eq:SaddlepointAppr}
\end{align}
The prime on the determinant denotes that the zero modes are excluded from the determinant.
The $\Delta \F$ in the exponent is defined as $\F[\cb]-\F[\phi_\rmi{meta}]$.
The negative eigenvalue, which is eliminated by the delta function in Eq.~\eqref{eq:GeneralArea}, reappears in the determinant due to the normalisation factor, $\mathcal{N}$.
The factor $\mc{V}$ denotes the contribution from zero modes, or deviations of the critical bubble which do not modify the effective free energy.
In three spatial dimensions, there are generically three zero modes due to the breaking of translation invariance, which give
\begin{align}
    \mc{V} &= V \prod_{i=1}^3\left[\int_{\mb{x}} (\partial_i \phi)^2\right]^{1/2} \\
    &=V \Delta\F^{3/2} \,, \label{eq:ZeroModes}
\end{align}
where $V$ is the volume of space, and on the second line we have assumed that the kinetic term of the field $\phi$ is canonically normalised.
There may be additional zero modes related to the breaking of internal symmetries, and their contribution can be solved for using the method of collective coordinates~\cite{Vainshtein:1981wh}.
In general their contribution is the volume of the zero-mode subspace of the configuration space.

The inclusion of the normalisation factor $\mathcal{N}$ in Eq.~\eqref{eq:GeneralArea}, and consequently of the negative eigenvalue in the determinant in Eq.~\eqref{eq:SaddlepointAppr}, means that $-\log\Sigma$ takes the form of the analytic continuation of an effective action~\cite{Langer:1967ax}.
As a consequence, for a gauge theory the gauge invariance of $\Sigma$ can be demonstrated order by order using the Nielsen-Fukuda-Kugo identities~\cite{Nielsen:1975fs,Fukuda:1975di}.

Finally, let us discuss the region of validity of the semiclassical results presented here.
As mentioned in the beginning of this section, there needs to be a hierarchy between the particle scattering and nucleation time scales, and this separation comes from the Boltzmann suppression of nucleating bubbles.
Thus, there are two ways for the expression to break down.
The first one is that the critical bubble is not exponentially suppressed by its effective free energy.
Physically, this corresponds to the bubble being comparable to a typical thermal fluctuation of the effective field.
Another possibility is for the fluctuation determinants to become exponentially large and to cancel the exponential suppression from the critical bubble.
This can happen in phase transitions which are weakly first order, i.e.\ close to being second order transitions, and was discussed in Ref.~\cite{Strumia:1998qq}.
Physically this corresponds to wildly distinct bubbles contributing to the transition -- an exponential variety of configurations.


\section{Effective theories for thermal bubble nucleation} \label{sec:formalism}

In this section, we present the EFT approach to calculating the statistical part of the bubble nucleation rate, $\Sigma$ in Eq.~\eqref{eq:divisionDynamicalStatistical}, which contains all the exponential contributions to the rate.
As discussed in the previous section, $\Sigma$ is the Boltzmann weighted transition surface area in the saddle-point approximation around the critical bubble.
The crucial step left out of classical nucleation theory is the derivation of the effective free energy from an underlying QFT, in essence to substantiate Langer's schematic Eq.~\eqref{eq:LangerFree}, repeated here:
\begin{equation}
    e^{-\beta \F[\phi]} = \sum_{\left[\substack{\mr{constrained}\\ \mr{microscopic\ variables}}\right]} e^{-\beta H} \,. \nonumber
\end{equation}
This is the step that we will focus on in Sec.~\ref{sec:formalism}.

We will outline the general principles of how to construct the effective free energy, from which can be derived all the quantities which enter $\Sigma$:
the Boltzmann factor, the transition surface and the critical bubble.
This will entail a general discussion of EFTs at high temperature, and of dimensional reduction, as well as a discussion of the specific features that are pertinent to bubble nucleation.
We will see how the use of EFT solves the double counting issue and uncontrolled derivative expansions.
Some relevant subtleties are discussed in appendices: the thin-wall regime in Appendix~\ref{appendix:thinwall} and lower temperature transitions in Appendix~\ref{appendix:lowerTemp}.

In thermal first-order phase transitions, the global minimum of the free energy shifts discontinuously with temperature.
Hence, their description requires taking thermal fluctuations into account.
In many thermodynamic computations, one integrates out all nonconstant thermal and quantum fluctuations to obtain the free energy for a flat, homogeneous background.%
\footnote{
    For a discussion of the interpretation of such (complex) effective potentials for homogeneous backgrounds see Ref.~\cite{Weinberg:1987vp}, and for further discussion of their inappropriateness for bubble nucleation by the same author see Ref.~\cite{Weinberg:1992ds}.
}
However, this cannot be the correct approach for describing bubble nucleation, because the configurations along the transition surface are not homogeneous configurations, but rather different bubbles.
Therefore, we need to leave some inhomogeneous modes unintegrated so that we can still describe the transition surface.
This leads to the schematic Eq.~\eqref{eq:LangerFree} for finding the effective free energy that describes nucleation.
There, enough of the fluctuations have been integrated out so that the effective free energy has the correct phase structure, but also enough have been left unintegrated so that it can still describe nucleating bubbles.

EFT is exactly the tool we need to implement Eq.~\eqref{eq:LangerFree}.
A top-down construction of an EFT amounts to mapping a full theory to a corresponding effective theory that can describe its IR behaviour, i.e.\ the long distance and low energy scales of the full theory.
The Lagrangian of the EFT should respect the unbroken symmetries of the full theory, and will be local if there exists a scale hierarchy.
The mapping can be carried out via matching the coefficients of the EFT Lagrangian so that its behaviour coincides with the IR behaviour of the full theory.
This matching effectively integrates out the higher energy scales, their contributions to IR physics manifest only through these coefficients.

\subsection{Scales of interest}

\begin{figure}
    \centering
    \includegraphics{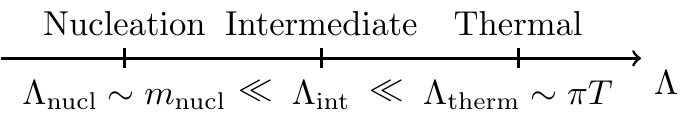}
    \caption{The energy scales, $\Lambda$, of particular interest in thermal bubble nucleation.
    Thermal and nucleation scales are always present and must be separated by a scale hierarchy, $\Lambda_{\mathrm{nucl}}\ll\Lambda_{\mathrm{therm}}$, for a local nucleation scale EFT to be able to capture the thermal corrections and for classicalisation to occur.
    The intermediate scale $\Lambda_{\mathrm{int}}$ arises naturally in some theories and can have a non-trivial effect on nucleation.
    Note that in the thin-wall regime $\LamNuc\sim m_{\text{nucl}}$ no longer holds, and is replaced by $\LamNuc\ll m_{\text{nucl}}$.}
    \label{fig:scales}
\end{figure}

In a first-order phase transition driven by the change of temperature, there are always at least two energy scales: the thermal scale, $\Lambda_{\mathrm{therm}}$, and the nucleation scale, $\LamNuc$ (\textit{cf.} Fig.~\ref{fig:scales}).
The former scale contains the thermal fluctuations and consequently the dominant thermal effects.
It is of order $\Lambda_{\mathrm{therm}}\sim \pi T$.
The latter scale contains the dynamics of nucleation.
Thus, it is the length scale of the bubble radii, which can be given in terms of the mass of the nucleating effective field, $\Lambda_{\mathrm{nucl}}\sim m_{\text{nucl}}$.
This point is discussed further around Eq.~\eqref{eq:idNuclScale}, where the identification is made.

Note, that the creation of the EFT for the nucleation scale, with the thermal contributions integrated out, is only possible if $\Lambda_{\mathrm{nucl}}\ll\Lambda_{\mathrm{therm}}$.
This is actually also the condition for applying the classical effective picture from Sec.~\ref{sec:background}, as in this case the nucleation scale modes are Bose enhanced.
The condition is quite natural in perturbative QFTs.
This is because the phase transition occurs when coupling-suppressed thermal corrections balance against tree-level terms.
For example, the effective mass squared is the sum of the tree-level mass squared and a term of the form (coupling)$\times T^2$.
Balancing these two terms yields a natural scale hierarchy (\textit{cf.} Eqs.~\eqref{eq:orderTildeMPhi1}, \eqref{eq:approx_Tc} below), with the temperature parametrically larger than the mass, as long as the theory is perturbative.
However, this reasoning can be violated, for example, when at zero temperature there are two phases with very similar vacuum energies, then only a small thermal correction is needed to cause a phase transition.
In this case, the EFT is out of reach.
Methods for handling non-high-temperature phase transitions with $\Lambda_{\mathrm{nucl}}\slashed{\ll}\Lambda_{\mathrm{therm}}$ are briefly discussed in Appendix~\ref{appendix:lowerTemp}.

An additional scale may arise between the thermal scale and the nucleation scale, which we will call the intermediate scale, $\Lambda_{\mathrm{int}}$ (\textit{cf.} Fig.~\ref{fig:scales}).
While some first-order phase transitions contain only the thermal and nucleation scales, an intermediate scale may be present and have a decisive effect on the dynamics of the transition.
Example of the latter are given in Secs.~\ref{sec:two_scalar} and \ref{sec:scaleshifters}.

Even if the intermediate scale contributions to nucleation are not of leading order, they can still give significant contributions to the masses of the nucleation scale, due to $\Lambda_{\mathrm{int}}$ enhancements.
These contributions are straightforward to quantify and account for using EFT.

The emergence of an intermediate scale may happen in theories with an intrinsic hierarchy of couplings or masses, through the resulting effects on the behaviour of thermal effective masses.
One possibility is that a field is relatively strongly coupled to a field undergoing a symmetry breaking transition.
This can happen, for example, in the cubic anisotropy model and gauge-Higgs models.
The cubic anisotropy model is studied in Sec.~\ref{sec:scaleshifters}.

Lastly, we want to note that in the thin-wall regime $\LamNuc\sim\mNuc$ no longer holds. This happens because the critical bubble radius becomes large due to the smallness of the free energy difference between the phases.
Hence, $\LamNuc\ll\mNuc$, and the scale of $\mNuc$ becomes an intermediate scale.

\subsection{Nucleation scale EFT from the imaginary-time formalism}

The statistical part of the nucleation rate, $\Sigma$ in Eq.~\eqref{eq:GeneralArea}, is a static equilibrium quantity.
Hence we may compute it using the imaginary-time formalism of QFT at finite temperature (\textit{cf.}\ Refs.~\cite{Matsubara:1955ws,Laine:2016hma}), which captures the equal time correlation functions of a theory at finite temperature.
Here, we will discuss the structure of the imaginary-time formalism relevant for obtaining the nucleation scale EFT.
This EFT will be identified with the effective description at the heart of classical nucleation theory; the link is elaborated in the next subsection.

The partition function in the imaginary-time formalism is given by
\begin{align} \label{eq:ImaginaryTimeZ}
    Z(T) &=\Tr[e^{-\beta \hat{H}}] \nonumber\\
    &= \int_{\mathrm{BCs}}\mathcal{D}\Phi\, \exp{-\int_0^\beta\dd\tau\int_{\mathbf{x}}\,\mathscr{L}_{\mathrm{E}}} \,,
\end{align}
where $\hat{H}$ is the Hamiltonian of the theory,
$\Phi$ denotes the full field content,
$\mathscr{L}_{\mathrm{E}}$ is the Euclidean Lagrangian,
and $\tau$ is called the Euclidean or imaginary time.
The physical equal time correlators are given by the equal imaginary-time correlators of this formalism.

The Euclidean time in Eq.~\eqref{eq:ImaginaryTimeZ} has finite extent equal to the inverse temperature, $\beta$, and the boundary conditions (BCs) are such that bosonic fields are periodic in the Euclidean time and fermionic fields are anti-periodic.
This gives rise to a discrete set of Fourier modes in the Euclidean time direction, the Matsubara frequencies: $\omega_n^{\rmi{b}}=2\pi Tn$ for bosons and $\omega_n^{\rmi{f}}=2\pi T(n+\frac{1}{2})$ for fermions.
We define the Matsubara mode decomposition as
\begin{equation}
    \Phi^{\rmi{b/f}}(\tau,\,\mathbf{x}) = \sum_{n \in \mathbb{Z}} \Phi^{\rmi{b/f}}_n(\mathbf{x})e^{i\omega_n^{\text{b/f}}\tau} \,,
\end{equation}
so that the dimensionalities of the Matsubara modes match those of the full fields.

The peculiar structure of the Euclidean time implies that the effective theory for energy scales much below the thermal scale (distance scales much longer than $\beta$) no longer contains the Euclidean time dimension, and hence is purely three dimensional.
This {\em high temperature dimensional reduction} was first studied in Refs.~\cite{Ginsparg:1980ef,Appelquist:1981vg,Nadkarni:1982kb,Landsman:1989be}
and was developed and systematised in Refs.~\cite{Farakos:1994kx,Kajantie:1995dw,Braaten:1995cm,Braaten:1995jr}, so facilitating higher order calculations.
It has since been used in a plethora of calculations -- for example, to calculate the thermal bubble nucleation rate on the lattice~\cite{Moore:2000jw,Moore:2001vf},
to calculate the phase diagram of the electroweak sector~\cite{Kajantie:1995kf,DOnofrio:2012phz}, and
to calculate the pressure of QCD to $\order{g^6\log g T^4}$~\cite{Kajantie:2002wa}.
For reviews, see Refs.~\cite{Kajantie:1995dw,Andersen:2004fp,Jakovac:2016zkg,Laine:2016hma,Ghiglieri:2020dpq}.

Dimensional reduction can be understood as creating an EFT for the light bosonic zero Matsubara modes, $\Phi^{\rmi{b}}_0(\mathbf{x})$, which are Bose enhanced.
It can be seen from the bare momentum space bosonic and fermionic propagators,
\begin{equation}
    \frac{1}{\mathbf{p}^2+(\omega_n^{\rmi{b}})^2+m_{\rmi{b}}^2}\,,\quad \frac{-i\slashed{P}+m_{\rmi{f}}}{\mathbf{p}^2+(\omega_n^{\rmi{f}})^2+m_{\rmi{f}}^2} \,,
\end{equation}
that the Matsubara frequencies yield an additive correction to the effective mass.
Thus, the bosonic nonzero Matsubara ($n\neq 0$) modes and all the fermionic modes become much heavier that the bosonic zero Matsubara mode, when the temperature is much larger than the mass of the bosonic field, $(\pi T)^2\gg m_{\rmii{b}}^2$.
The zero Matsubara mode is independent of the Euclidean time, and thus, this dimension is absent in the EFT.
The EFT contains only the three physical, spatial dimensions.

The partition function of the full theory is related to that of the EFT by~\cite{Braaten:1995cm,Braaten:1995jr}
\begin{equation}\label{eq:DRPartFunc}
    Z(T) = e^{-V\beta f_{\rmii{therm}}(\Lambda)}\int^{(\Lambda)}\mathcal{D}\phi\, \exp{-\int_{\mathbf{x}}\,\mathscr{L}_{\mathrm{eff}}} \,,
\end{equation}
where $\Lambda$ is the factorisation (or the matching) scale, which separates the thermal scale from the lower scales and acts as the UV cut-off for the dimensionally reduced theory.
The field-independent factor in Eq.~\eqref{eq:DRPartFunc}, which has been pulled out of the path integral, is the free energy contribution purely from the thermal scale, $f_{\rmii{therm}}(\Lambda)$.
It will cancel from the nucleation rate, as this only depends on the difference between the action evaluated on the bubble and on the metastable phase (\textit{cf}. Eq.~\eqref{eq:SaddlepointAppr}), so $f_{\rmii{therm}}(\Lambda)$ is irrelevant for us.
Here, $\phi$ denotes all the light (bosonic) fields of the EFT, which correspond to the zero-Matsubara modes of the full theory with $m^2_{\rmi{b}}\ll (\pi T)^2$.

Let us assume for now that only the thermal and nucleation scales are present.
In this case, the steps from Eqs.~\eqref{eq:ImaginaryTimeZ} to \eqref{eq:DRPartFunc} accomplish the schematic coarse-graining of Eq.~\eqref{eq:LangerFree}.
This yields a path integral for the Bose-enhanced modes of the nucleation scale, which corresponds precisely with the classical path integral structure of Eq.~\eqref{eq:GeneralArea}.
Thus, we can identify
\begin{equation} \label{eq:snucl}
\beta\F \equiv \sn = \int_{\mathbf{x}}\,\mathscr{L}_{\mathrm{eff}} \,,
\end{equation}
forming the bridge from QFT to classical nucleation theory.
With this identification, we can utilise the analysis of Sec.~\ref{sec:background} to compute the statistical part of the nucleation rate to one-loop accuracy within the EFT, Eq.~\eqref{eq:SaddlepointAppr}.

In theories with additional intermediate scales, the partition function for the nucleation scale EFT still has the general structure of that given in Eq.~\eqref{eq:DRPartFunc}.
One only has to lower the factorisation scale, $\Lambda$, so that it separates the nucleation scale, $\LamNuc\sim\mNuc$, from all higher scales.
This can be carried out by creating a chain of EFTs, with each link matched to its neighbours.
Note that the nucleating DoF remains dynamical in the nucleation scale EFT due to $\LamNuc\sim\mNuc$.

Creating the chain of EFTs, from the thermal down to the nucleation scale, organises powers of ratios of scales, and permits one to construct a description of the nucleation scale which is accurate up to corrections of known magnitude, and which is systematically improvable.
We will show an example with one scale hierarchy in Sec.~\ref{sec:one_scalar}, examples with two scale hierarchies in Sec.~\ref{sec:two_scalar} and \ref{sec:scaleshifters}, and the thin-wall regime, where there is an additional hierarchy $\LamNuc\ll\mNuc$, in Appendix~\ref{appendix:thinwall}.

In this EFT framework, we can now see how the catch-22 discussed in the Introduction is resolved.
The factorisation scale, $\Lambda$, divides the fluctuations into two groups, based on their energy $E$, which are respectively counted only once,
\begin{equation} \label{eq:SigmaModes}
\Sigma = \underbrace{\mc{V} \sqrt{\abs{\frac{\det (\sn''[\phi_\rmi{meta}]/2\pi)}{\det^\prime (\sn''[\cb]/2\pi)}}}}_{\text{modes }E<\Lambda} \, \underbrace{e^{-\sn[\cb]}}_{\text{modes }E>\Lambda} \,.
\end{equation}
First, the fluctuations with energies higher than $\Lambda$ are integrated out, yielding the effective action of the nucleation scale, $\sn$.
The critical bubble is then solved for, as the saddle point of $\sn$ (Eqs.~\eqref{eq:StationaryPoint} and \eqref{eq:snucl}).
Finally, the fluctuations with energies lower than $\Lambda$ are integrated over, in the background of the critical bubble, resulting in the fluctuation determinant.
This can be identified as the contributions from differently shaped bubbles.%
\footnote{
    The negative eigenvalue in $\Sigma$ does not arise as a result of integrating over the negative eigenmode via analytical continuation, as is the case in the vacuum decay rate.
    Here, it is simply included as a conventional normalisation (\textit{cf.} Eq.~\eqref{eq:GeneralArea}).
    Hence, the determinants can truly be identified with contributions from differently shaped nucleating bubbles, up to this normalisation.
}
Altogether, this process avoids uncontrolled gradient expansions (\textit{cf.} Sec.~3.5 of Ref.~\cite{Croon:2020cgk}) by making use of the natural hierarchies of scale.
Further, all quantities are completely real, there are no stray imaginary parts, because fluctuations below $\Lambda$ are not integrated out before solving for the critical bubble.

Finally, we would like to make a remark concerning loop orders and the construction of the nucleation scale EFT.
The classical formalism of Sec.~\ref{sec:background}, and crucially the factorisation of the nucleation rate into statistical and dynamical parts (Eq.~\eqref{eq:divisionDynamicalStatistical}), is based upon a one-loop, saddle-point approximation within the nucleation scale EFT, identified as the classical effective description.
This means that the calculation of higher loop corrections to the nucleation rate coming from {\em within} the nucleation scale EFT goes beyond the scope of this article.
On the contrary, there is no impediment to incorporating higher loop corrections from the higher energy scales into the nucleation scale EFT, which then gives the classical description.
The construction only relies upon general principles of EFT.
This is important because, at low loop order, the higher energy scales typically give parametrically larger contributions.
Secs.~\ref{sec:two_scalar} and \ref{sec:scaleshifters} provide examples at two-loop order.

\subsection{Subtleties of matching classical nucleation theory}

The EFT framework resolves many complications in the calculation of the thermal nucleation rate, and offers an intuitive physical picture of the nucleating bubbles.
However, the framework is built on an implicit assumption: that the structure of the transition surface (the background for the bubbles) is dominated by contributions due to the higher energy scales, and not the nucleation scale nor a lower energy scale.
This is requisite for a local description of the transition surface.
In the following, we discuss the range of validity of this assumption, and related subtleties.

To set the scene, we will suggest a suitable definition of a local description, and discuss its relationship with EFT.
Following this, we will consider the effects of lower energy scale fluctuations on the critical bubble.
These will reveal whether a local description is possible.
Finally, we will discuss some difficult cases and ways to proceed in them.

By a local description of the transition surface, we mean that there exists a local free energy functional (Eq.~\eqref{eq:LangerFree}), containing a finite number of spatial derivatives, whose extremum gives a systematically improvable approximation for the critical bubble (Eq.~\eqref{eq:StationaryPoint}).
The requirement of systematic improvability is important.
It means that by including higher-order terms, perhaps involving a larger, but finite, number of spatial derivatives, one can systematically improve the approximation to the critical bubble, with higher-order terms being successively smaller.

The EFT approach is built upon the derivative expansion: by integrating out only short-wavelength modes, higher derivative terms in the effective description are small.
Going beyond the EFT approach and integrating out any modes from the nucleation scale or lower inevitably violates the derivative expansion, yielding a nonlocal description. Physically, this is because the integrated fluctuations take place on the same length scales as the nucleating bubbles.
Consequently, if a local description of the transition surface exists, the nucleation scale EFT implements it.

In the EFT approach, the critical bubble lives in the background of shorter wavelength fluctuations; it is determined by the tree-level action of the EFT.
The (nonlocal) effects on the critical bubble due to fluctuations within the EFT are not included.
These take the form of loop-level tadpole diagrams (see for example Ref.~\cite{Fukuda:1975di}), induced by integration over the DoFs of the EFT.
Being loop-induced, these should be small corrections to the tree-level terms, as long as the nucleation scale EFT is perturbative.

Diagrammatically, the correction to the critical bubble from a DoF in the nucleation scale EFT can be understood as follows.
At one-loop level, integration over the DoF results in a fluctuation determinant represented as
\begin{equation}\label{eq:shapechangeonelooporigins}
    \feynalign{\oneloopbubbledotted} \,.
\end{equation}
At LO, this affects the critical bubble configuration through
\begin{equation}\label{eq:shapechangesaddle}
    \feynalign{\shapechangesaddle} \,,
\end{equation}
where the solid line is the propagator of the nucleating DoF.
This correction to the critical bubble by itself leads to the following contribution to the exponent of the statistical part of the nucleation rate:
\begin{equation}\label{eq:shapechangecontrib}
    \feynalign{\shapechangenucl} \,.
\end{equation}

This two-loop dumbbell diagram must be small compared to the tree-level action terms, if the nucleation scale EFT is to give a local description for the critical bubble, admitting a saddle-point approximation.
The dumbbell is naturally small if the nucleation scale EFT is perturbative, being suppressed by the loop-expansion parameter of the EFT.
In this case, so too are higher-order corrections to the critical bubble, which arise at higher loop orders.

One should be mindful also to account for potential enhancements by ratios of scales, yet these only serve to reinforce the validity of the EFT approach.
If the DoF of the dotted lines is lighter than the nucleation scale, then the dumbbell diagram is even further suppressed than its loop counting would suggest.
On the other hand, the effects of heavier DoFs, which could potentially give large corrections, have already been accounted for through the construction of the nucleation scale EFT.

The above considerations leave only one possibility for the nucleation scale fluctuations to give LO contributions to the critical bubble, and hence for a nonlocal description to be necessary: the loop expansion within the EFT must break down.
This can happen either due to strong coupling, or due to a sufficiently large number of weakly-coupled degrees of freedom, both of which cases we discuss below.
We also discuss two other subtle cases, for which local descriptions are nevertheless possible: the thin-wall regime, and models in which the mass of a field changes so drastically that it is associated with parametrically different scales on different parts of the critical bubble.

When the nucleating DoF is strongly coupled, the loop expansion breaks down altogether, and therefore so too does the tadpole expansion.
In three dimensions, the loop expansion parameter contains an inverse power of the field mass.
Hence, the self-coupling of the nucleating field increases if the transition approaches a spinodal decomposition, at which point the mass vanishes.
Too near to the spinodal decomposition, or in too weak a first-order transition, the only possibility is thus to resort to nonperturbative lattice Monte-Carlo simulations.

If the nucleation scale EFT contains a sufficiently large number $N$ of weakly coupled fields, the dumbbell term, which scales as $N^2$, can become of LO.
In this case however, some higher-loop diagrams are also of LO, such as the triple dumbbell at three-loop order, showing the breakdown of the vanilla loop expansion.
As before, this case can be studied on the lattice.
Another possibility, if the LO $N$ dependence can be resummed to produce a consistent perturbative expansion, is to adopt the iterative nonlocal approach~\cite{Surig:1997ne,Garbrecht:2015yza,Ai:2018guc,Ai:2020sru}.
In this case, the construction of the nucleation scale EFT still plays an important role, as it accounts for the higher energy scales, and helps to give a suitable starting point for the iterative procedure.

In the thin-wall regime, there are near-perfect cancellations amongst the tree-level terms such that the free energy difference between the phases is anomalously small.
Therefore, deep enough in the thin-wall regime, loop-level contributions from the scale of $\mNuc$ can become of the same order as the tree-level free energy difference.
Consequently, these loop-level contributions cannot be treated perturbatively in the background of the usual tree-level critical bubble.
However, the nucleation scale is now its own IR scale, $\LamNuc\sim 1/R\ll\mNuc$, where $R$ is the bubble radius.
Thus, one can integrate out the now-intermediate scale of $\mNuc$ to obtain a description valid in the thin-wall regime.
This is discussed in Appendix \ref{appendix:thinwall}.

Lastly, we consider the case where the the effective mass of a field shifts so drastically between the phases that the scale to which it is associated changes, being much heavier than the nucleation scale in at least one part.
We refer to these kinds of fields as {\em scale-shifters}.
Seemingly, they are difficult to accommodate within EFT, because they are associated with different scales on different parts of the critical bubble.
Cosmologically perhaps the most relevant case, is that in which a light field jumps to the intermediate scale in a symmetry breaking transition.
This kind of behaviour is, for example, exhibited in gauge-Higgs theories by the spatial components of the gauge field.
In Sec.~\ref{sec:scaleshifters}, we show how this situation can be handled perturbatively, while keeping the residual errors clear.
In fact, the errors can even be eliminated entirely from the exponential order with techniques for evaluating fluctuation determinants or with the iterative nonlocal method.
In a nutshell, the field should be associated with the intermediate scale, and integrated out into the nucleation scale effective action.
This is because the tadpoles from the intermediate scale outweigh the nonlocal errors from the nucleation or a lower scale.

\subsection{Nucleation scale EFT construction} \label{sec:eftConstruction}

Finally, we would like to discuss the process of constructing the EFT of the nucleation scale.
This proceeds in the same general way as in other contexts:
the low energy degrees of freedom are identified,
the most general action containing these degrees of freedom and obeying the relevant symmetries is written down, truncated at some order,
and the low energy behaviour of the full theory is matched against that of the EFT.
For a general review of EFT see for example Ref.~\cite{Manohar:2018aog}, and for reviews of high-temperature dimensional reduction see Refs.~\cite{Andersen:2004fp,Jakovac:2016zkg,Laine:2016hma,Ghiglieri:2020dpq}.
For the case at hand, we will underline some pertinent aspects: power counting and the truncation of the EFT, and we will briefly touch on the practicalities of matching.
Such practicalities are also demonstrated in the examples of Secs.~\ref{sec:one_scalar}, \ref{sec:two_scalar} and \ref{sec:scaleshifters}.

In the presence of a hierarchy of scales, the counting of loop orders is no longer a reliable way to organise perturbative calculations.
This is because, in addition to the couplings, the ratio of scales provides a new small parameter which may enhance or suppress infinite classes of Feynman diagrams.
EFT provides a systematic approach to carrying out the necessary resummations of diagrams, by means of a dual expansion in powers of couplings and in the ratios of energy scales $\Lambda_{\mathrm{low}}/\Lambda_{\mathrm{high}}$.
If the ratio of scales can be related to the couplings, such as in high-temperature dimensional reduction, the result is a pure expansion in powers of couplings, though one that differs from the loop expansion.

In the absence of a computation of the dynamical part of the nucleation rate, there is a baseline error which leads to a natural truncation of the EFT.
The dynamical part enters as a prefactor to the nucleation rate (\textit{cf.} Eq.~\eqref{eq:divisionDynamicalStatistical}), and it is not expected to be exponentially large in the power counting.
Hence, this baseline error is an $\order{1}$ (or logarithmic) correction compared to the nucleation action, and one need only keep contributions which are parametrically larger than this.
Of course, if one calculates the dynamical part of the nucleation rate, the action of the EFT will have to be calculated to the corresponding accuracy.

To be able to truncate the EFT action appropriately, one must be able to estimate the magnitude of different contributions to the bubble action.
To do so one first accounts for the higher scales at leading order (LO), from which one can deduce the scalings of the critical bubble.
Next-to-leading order (NLO) and successive higher-order terms in the EFT action can then be estimated by evaluating them on the critical bubble.
Breaking this down, one needs estimations of the bubble volume, $V_{\rmii{CB}}$, the field value at the centre of the bubble, $\phi_0\equiv \cb(r=0)$, and the gradient, $\grad$, evaluated on the bubble.
The contribution from a given action term can then be estimated as $V_{\rmii{CB}}\times\text{Lagrangian term}$.

The gradient $\grad$, evaluated on the critical bubble, can be linked to the mass of the nucleating field, $m_{\text{nucl}}$.
The critical bubble is a localised field configuration, and a saddle point of the action of the nucleation scale EFT.
As such, the kinetic and potential terms in the Lagrangian are of the same order of magnitude when evaluated on the critical bubble,
leading to $(\grad{\cb})^2\sim m_{\text{nucl}}^2\cb^2$, or simply
\begin{equation} \label{eq:gradientSimMass}
    \grad\sim m_{\text{nucl}}
\end{equation}
when evaluated on the bubble.
Eq.~\eqref{eq:gradientSimMass} can be demonstrated with scaling arguments akin to those used to prove the virial theorem, or Derrick's theorem~\cite{Derrick:1964ww}.

From Eq.~\eqref{eq:gradientSimMass} we directly obtain the length scale at which the critical bubble configuration varies. Thus obtaining the nucleation scale,
\begin{equation}\label{eq:idNuclScale}
    \LamNuc\sim\mNuc \,,
\end{equation}
a central result for constructing the nucleation scale EFT.
This holds as long as the critical bubble is not thin-walled.
In the thin-wall regime, the free energy difference between phases is anomalously small, resulting in the emergence of a new infrared scale in the EFT, associated with the bubble radius.
There, Eq.~\eqref{eq:idNuclScale} no longer holds, and the nucleation scale satisfies $\LamNuc\ll\mNuc$.

Now, we can also estimate the volume of the critical bubble as
\begin{equation}\label{eq:BubbleVolume}
    V_{\rmii{CB}}\sim m_{\text{nucl}}^{-3} \,.
\end{equation}
From this one finds that to calculate all parametrically large contributions to $\sn$ requires calculating the Lagrangian of the EFT up to but not including $\order{m_{\text{nucl}}^3}$.

Finally, there is the estimation of $\phi_0$.
This can be found by equating the order of magnitude of the mass term in the potential with the term responsible for making the potential go down again towards the stable minimum, akin to the reasoning for obtaining Eq.~\eqref{eq:gradientSimMass}.
Near the critical temperature, this implies that $\phi_0\sim\phi_{\mathrm{stable}}$.
However, if there is significant supercooling it may be that $\phi_0\ll\phi_{\mathrm{stable}}$.

The estimation of the field variable is important for determining the relevance of an operator, but it can also reveal a need for resummation:
In Examples 2 and 3, the field value changes so much that
its effect on the mass of the integrated-out field is of LO.
Thus, the change has to be resummed into the mass.
Conversely this is not necessary in Example 1.

Above, we have considered the magnitude of terms evaluated on the critical bubble, in order to understand how to truncate the effective action.
One might still be concerned about how various terms affect the critical bubble itself, through corresponding tadpole diagrams.
This however is not a problem, as we shall now demonstrate.

Let us represent some particular contribution to the effective action by
\begin{align}\label{eq:actionleftout}
\Delta S = \int_{{\bf x}} \Delta \ms{L} \equiv \feynalign{\blob} \,,
\end{align}
which we will assume to be parametrically smaller than LO.
This term affects the critical bubble, which in turn leads to the following additional correction to the exponent of $\Sigma$:
\begin{align}\label{eq:actionleftoutshape}
\feynalign{\blobshape} \,.
\end{align}
Assuming for simplicity that we are not in the thin-wall regime, then the internal line must be of the nucleation scale as there are no other scales present.
Thus, we can power count the ratio of Eqs.~\eqref{eq:actionleftoutshape} and \eqref{eq:actionleftout} to be
\begin{align}
    \feynalign{\blobshape}\bigg/\feynalign{\blob}\;&\sim\;\frac{(\Delta\ms{L}')^2}{\mNuc^2 \Delta\ms{L}} \\
    &\sim\;\frac{\Delta\ms{L}}{\mNuc^2\phi^2} \,.\label{eq:dumbbellsmall}
\end{align}
This ratio is small as long as the term $\Delta \ms{L}$ is smaller than the LO mass term, i.e.\ as long as $\Delta \ms{L} \ll \ms{L}^{\rmii{(LO)}}$.
Hence, compared to the LO, the effects from the change of the critical bubble are of second order in the small expansion parameter.
This is, in essence, an elementary result of second-order perturbation theory.

Using similar power counting arguments as above, one can truncate the differential equation describing the critical bubble, Eq.~\eqref{eq:StationaryPoint}, to be more amenable.
Dropping a term from the action, Eq.~\eqref{eq:actionleftout}, gives rise to a corresponding tadpole expansion, whose leading order is the dumbbell diagram, Eq.~\eqref{eq:actionleftoutshape}.
The expansion is perturbative if the dropped term is of subleading order.
Conversely, including a term into the differential equation describing the critical bubble effectively resums the corresponding tadpole expansion to all orders.

So, the above estimates allow one to rank the operators of the EFT, and to truncate them, in order to achieve an appropriate accuracy for the nucleation rate.
Once this is complete, the next task is to match the coefficients of these operators.

The coefficients of the EFT can be chosen in order
to match the corresponding long distance correlation functions of the theories.
The task is immensely simplified by treating all IR-quantities (masses of the EFT-fields and external momenta) via {\em strict perturbation theory} and utilising dimensional regularisation~\cite{Braaten:1995cm,Braaten:1995jr}; for a review see Sec.~5.7 of Ref.~\cite{Manohar:2018aog}.
In this approach, on the EFT-side all loop integrals vanish due to being scale free.
Note that the renormalisation scale of the full theory must be run down to the EFT cut-off (or matching scale) to treat both sides equally.
The result is essentially to add the 1-EFT-particle-irreducible diagrams, whose EFT-field propagators are IR-regulated with dimensional regularisation, to the corresponding couplings of the EFT.%
\footnote{
    While it is often stated that one needs only to match the 1PI correlation functions~\cite{Kajantie:1995dw,Hirvonen:2020jud}, for example in the presence of heavy scalar particles there can also be contributions from diagrams containing reducible internal lines of these scalars~\cite{Brauner:2016fla,Manohar:2020nzp}.
}
Note that it is conventional to scale the EFT fields to have canonical kinetic terms.
This scaling enters the matching relations;
see for example the $\sqrt{T}$ in the matching relation of Eq.~\eqref{eq:tadpole_diagrams_example1}.

Intuitively, the EFT description shrinks to points the short distance parts of long distance correlation functions.
As an illustrative example from Ref.~\cite{Laine:2016hma}, consider a two-loop sunset diagram where the external legs are of a light field, and internally there is an additional heavy field.
The diagram can be split into two parts, depending on whether the momentum of the light field internal line is above or below the EFT cut-off:
\begin{align}
\feynalign{\sunset{plain}{dashed}} &= 
\; \feynalign{\sunsetpieceone{plain}} \;
+ \;  \feynalign{\sunsetpiecetwo{plain}} \;
.
\end{align}
Here the full lines refer to the light field, the dashed lines to the heavy field and the black circles refer to the two terms in the effective description.
The first term accounts for the matching of 2-point functions, whereas the latter comes from the matching of 4-point functions and then computing the loop-diagram within the EFT (consider joining two external legs in the middle diagram of Eq.~\eqref{eq:lam3_chi_phi_matching}).


\section{Example 1: One scale hierarchy} \label{sec:one_scalar}

The approach outlined in Sec.~\ref{sec:formalism} is model independent, relying only on the existence of a hierarchy of scales.
To put the formalism in context however, we apply it to three theories, starting with the simplest case where there is only one hierarchy of scales, that between the thermal and the nucleation scales, $\LamTh \gg \LamNuc$.

We first consider the simplest possible theory which may display a first-order phase transition, that of a real scalar field.
When this theory has a $Z_2$-symmetry, the thermal phase transition in known to be of second order~\cite{ZinnJustin:2002ru,Sun:2002cc}, so we must include $Z_2$-breaking terms.
In four dimensions, the most general renormalisable theory of a real scalar field, $\Phi$, is the following
\begin{align}
    \ms{L}_{\Phi} &=
    \frac{1}{2}\left(\partial_\mu \Phi\right)^2
    + V(\Phi) \,, \\
    V(\Phi) &=
    s \Phi
    + \frac{1}{2}m^2\Phi^2
    + \frac{1}{3!}g\Phi^3
    + \frac{1}{4!}\lambda \Phi^4 \,.
\end{align}
In addition, we add a Dirac fermion, $\Psi$, coupled to the scalar field via the Yukawa interaction,
\begin{equation}
    \ms{L}_{\Psi} + \ms{L}_{\rmi{Yukawa}} =
    \bar{\Psi}\left(\slashed{\partial} + m_\Psi\right)\Psi
    + y\Phi \bar{\Psi}\Psi \,,
\end{equation}
so that the total (Euclidean) Lagrangian of our first example takes the form,
\begin{equation}
    \ms{L} = \ms{L}_{\Phi} + \ms{L}_{\Psi} + \ms{L}_{\rmi{Yukawa}} \,.
\end{equation}
For simplicity we will make the natural assumption that the couplings of the theory are all equally perturbative, $\lambda \sim g^2/m^2 \sim y^2$, and that there are no mass hierarchies in the theory at zero temperature, $m \sim m_\Psi$.
In this case the loop expansion and the couplings expansion agree at zero temperature.
Further, for there to be a phase transition, thermal fluctuations must modify the effective potential of the theory at LO.
The leading effect of thermal fluctuations amounts to a shift in $m^2$ of order $\lambda T^2$ and a shift in $s$ of order $g T^2$.
As a consequence, in the vicinity of the phase transition one expects $m^2\sim \lambda T^2$ and $s\sim g T^2$.
These relations define our power counting for this theory.

The modes of the thermal scale are integrated out using dimensional reduction, so constructing the EFT which describes the nucleation scale.
The thermal modes are the entire field $\Psi$ as well as the nonzero Matsubara modes of $\Phi$.
The 3d EFT is described by the most general Lagrangian for the remaining degree of freedom, a single real scalar field,
\begin{align}
    \ms{L}_3 &= 
    \frac{1}{2}\left(\grad \phi\right)^2
    + V_3(\phi) \,,
    \\ \label{eq:V3_example1}
    V_3(\phi) &=
    s_3 \phi
    + \frac{1}{2}m_3^2\phi^2
    + \frac{1}{3!}g_3\phi^3
    + \frac{1}{4!}\lambda_3 \phi^4 \,.
\end{align}

The construction of the 3d EFT is standard.
As outlined in Sec.~\ref{sec:eftConstruction}, it proceeds via matching static infrared quantities calculated in both the full theory and the EFT, so determining the effective field $\phi$ and the effective couplings of Eq.~\eqref{eq:V3_example1}.
These are matched order-by-order in powers of couplings.
Calculations of the static infrared quantities take place with a cutoff, or matching scale, $\Lambda$, which acts as an infrared cutoff for the full theory, as well as an ultraviolet cutoff for the EFT.
In the end, dependence on $\Lambda$ should drop out of all physical quantities.
The most economical method is to match 1-$\Phi/\phi$-irreducible correlation functions, and to use dimensional regularisation to introduce the cutoff.
For this theory, 1-$\Phi/\phi$-irreducible is equivalent to 1-particle-irreducible, as there are no fermion tadpole diagrams (fermions do not take a nonzero expectation value).
Further simplifications follow from utilising the strict perturbation theory of Refs.~\cite{Braaten:1995cm,Braaten:1995jr}, in which one treats IR masses, external momenta and tadpoles as perturbations.

To truncate the EFT appropriately, we must first determine the parametric magnitude of the nucleation scale mass, $m_{\text{nucl}}^2=V_3''(\phi_{\text{meta}})$, and the value of the field at the centre of the critical bubble $\phi_0$.
Shifting to a field basis where $s_3=0$, and equating the orders of magnitude of the remaining terms in the potential of the effective theory,
\begin{equation}\label{eq:parametricsNearTcEx1}
\mNuc^2\phi_0^2 \sim g_3 \phi_0 ^3 \sim \lambda_3 \phi_0^4 \,,
\end{equation}
results in
\begin{equation} \label{eq:orderTildeMPhi1}
m_{\text{nucl}}^2 \sim \lambda T^2, \qquad \phi_0 \sim \sqrt{T} \,.
\end{equation}
Note that, these relations can change if there is parametrically large supercooling, because the second relation in Eq.~\eqref{eq:parametricsNearTcEx1} holds only in the vicinity of the critical temperature.

The bubble volume is $m_{\text{nucl}}^{-3}\sim \lambda^{-3/2}T^{-3}$, and consequently the nucleation action is $m_{\text{nucl}}^{-1}\phi_0^2\sim \lambda^{-1/2}$ at LO.
Hence, to describe the nucleation action at $\order{1}$ accuracy, i.e.\ exponential order for the nucleation rate, requires contributions to the nucleation scale effective Lagrangian at $\mc{O}(\lambda T^3)$.

To construct the EFT Lagrangian at $\mc{O}(\lambda T^3)$ accuracy, the effective three-point and four-point couplings need only be matched at tree-level.
The same is true for the matching of the effective field, $\phi$: one-loop corrections resulting from matching the $\order{p^2}$ momentum dependence of the two-point functions are of higher order.
The only loop-corrected matching relations at this order are for the tadpole and mass terms, which are corrected at one-loop order.

Subleading corrections from the thermal scale correct the nucleation scale effective Lagrangian at $\order{\lambda^2T^3}$, and hence correct the nucleation action at $\mc{O}(\sqrt{\lambda})$.
Since a calculation of the nucleation rate at this accuracy would require calculating the dynamical part and also the two-loop corrections from the EFT field, which are beyond our scope, these subexponential corrections wind up in the uncertainties of the final results of this section, Eqs.~\eqref{eq:result_1} and \eqref{eq:parametric_result_1}.

Diagrammatically the matching relation for the tadpole reads
\begin{align} \label{eq:tadpole_diagrams_example1}
-s_3\sqrt{T} &= 
\; \feynalign{\tadpoletree{plain}{crossed dot}}\; 
+ \; \feynalign{\tadpole{plain}}\; 
+ \; \feynalign{\tadpoleinsertion{plain}{fermion}{crossed dot}}
 \\
&= -s
- \frac{g}{2} \sumint{P} \frac{1}{P^2}
+ y m_\Psi \sumint{\{P\}} \frac{\mr{tr}(\slashed{P}\slashed{P})}{P^4} \,,
\label{eq:tadpole_loop_integrals}
\end{align}
where plain lines denote the scalar, lines with arrows denote the fermion, and crossed dots denote tree-level insertions, of the scalar tadpole and the fermion mass.
The results for each diagram are shown below, appearing in the same order.
Our notation for momenta and loop integration are standard~\cite{Braaten:1995cm,Braaten:1995jr} and are given in Appendix~\ref{appendix:notation}.

The diagrammatic matching for the mass reads
\begin{align} \label{eq:mass_diagrams_example1}
-m_3^2 &= 
\; \feynalign{\selfenergytree{plain}{crossed dot}} \;
+ \;  \feynalign{\selfenergyfour{plain}{plain}} \;
+ \; \feynalign{\selfenergythree{plain}{fermion}}
 \\
&= -m^2
- \frac{\lambda}{2} \sumint{P} \frac{1}{P^2}
+ y^2 \sumint{\{P\}} \frac{\mr{tr}(\slashed{P}\slashed{P})}{P^4} \,,
\end{align}
where the crossed dot denotes a tree-level scalar mass insertion and we have dropped all contributions smaller than $\mc{O}(\lambda T^2)$.
The right hand sides of both Eqs.~\eqref{eq:tadpole_diagrams_example1} and \eqref{eq:mass_diagrams_example1} are evaluated in the full theory at high temperature, treating the particles as massless within loop sum-integrals due to the strict perturbation expansion in the IR quantities~\cite{Braaten:1995cm,Braaten:1995jr}.

The relevant one-loop sum-integrals are known analytically, and can be found, for example, in Appendix~A of Ref.~\cite{Braaten:1995jr}.
Inserting the integrals gives the matching relations at this order:
\begin{align}
    \phi(\mb{x}) &= \frac{1}{\sqrt{T}}\Phi_0(\mb{x}) \,, \\
    s_3 &= \frac{s}{\sqrt{T}} + \frac{T^{3/2}}{24}\left(g + 4 y m_{\Psi }\right)\label{eq:DRphi1}
    \,, \\
    m_3^2 &= m^2 + \frac{ T^2}{24}\left(\lambda + 4 y^2\right)
    \,, \\
    g_3 &= \sqrt{T} g
    \,, \\
    \lambda_3 &= T \lambda \,.
\end{align}
This completes the construction of the effective description of the nucleation scale,
\begin{equation} \label{eq:sn_example1}
    \sn = \int_{\mb{x}}\ \ms{L}_3 \,.
\end{equation}

Note that, unlike the effective potential of the full theory which is commonly utilised in bubble nucleation calculations, the potential of the nucleation scale EFT is real for all $\phi$.
It is also independent of the renormalisation scale $\Lambda$ up to the order we have calculated.

The equilibrium thermodynamics of this EFT has been studied in Ref.~\cite{Gould:2021dzl}.
To understand the phase diagram it is convenient to shift the field by a temperature dependent constant, $\phi \to \phi - g_3/\lambda_3$, after which the tree-level potential takes the form
\begin{align} \label{eq:V_g30}
V_3(\phi) &=
\bar{s}_3\phi
+ \frac{1}{2} \bar{m}_3^2\phi^2
+ \frac{1}{4!} \lambda _3 \phi^4
\,,
\end{align}
up to an irrelevant overall constant.
We have introduced overlines to denote the parameters in this shifted field basis,
\begin{align}
    \bar{s}_3 &= s_3 + \frac{g_3^3}{3 \lambda _3^2} - \frac{g_3 m_3^2}{\lambda _3} \,, \\
    \bar{m}_3^2 &= m_3^2-\frac{g_3^2}{2 \lambda _3} \,.
\end{align}
Eq.~\eqref{eq:V_g30} shows that the general theory reduces to the $Z_2$-symmetric theory in the presence of an external field $\bar{s}_3$.
If $\bar{s}_3$ changes sign as a function of temperature, there is a phase transition, and (at tree-level) this transition is first order if $\bar{m}_3^2 < 0$.
Beyond tree-level the endpoint of the line of first-order transitions shifts slightly away from zero, but the overall picture of the phase diagram is unchanged.
In the special case of $s=g=m_\Psi=0$, the 3d EFT has a $Z_2$-symmetry and the transition is of second order.

Being interested in bubble nucleation, let us consider a point in the parameter space of this 4d model in which there is a first-order phase transition.
With the effective action of the nucleation scale in hand, the statistical part of the bubble nucleation rate can be constructed using standard methods.
In particular, the critical bubble is the appropriate saddle point of the effective action; see Eq.~\eqref{eq:StationaryPoint}.
It is O(3)-symmetric~\cite{Blum:2016ipp}, and hence is a function only of the radial coordinate $r$, satisfying
\begin{equation}
   \frac{d^2\cb(r)}{dr^2} + \frac{2}{r}\frac{d\cb(r)}{dr} = V_3'(\cb(r)) \,.
\end{equation}
The boundary conditions are such that the field is regular at $r=0$, and approaches the metastable phase as $r\to\infty$.

To complete the saddle-point approximation of the statistical part of the nucleation rate requires calculating the functional determinant of fluctuations in the background of the critical bubble; see Eq.~\eqref{eq:SaddlepointAppr}.
Numerical methods for doing this are discussed briefly in Appendix~\ref{appendix:numerics}.

Putting it all together, the final result for the statistical part of the nucleation rate reads
\begin{align} 
    \Sigma &= 
    V\sn[\cb]^{3/2}\sqrt{\abs{\frac{\det  (\sn''[\phi_\rmi{meta}]/2\pi)}{\det^\prime (\sn''[\cb]/2\pi)}}}\, e^{-\sn[\cb]} \nonumber \\
    &\quad
    \times \left(1 + \mc{O}(\sqrt{\lambda})\right) \,.
    \label{eq:result_1}
\end{align}
Parametrically, this result is of the form
\begin{align} 
    \Sigma &= 
    a\lambda^{3/4} T^3 V \exp\left(-\frac{b}{\sqrt{\lambda}}\right)
    \left(1 + \mc{O}(\sqrt{\lambda})\right) \,,
    \label{eq:parametric_result_1}
\end{align}
where $a$ and $b$ are $\order{1}$ temperature-dependent values, calculable using the approach presented, and we have used that the low-lying eigenvalues of $\sn''$ scale as $m_{\text{nucl}}^2$.

The $\mc{O}(\sqrt{\lambda})$ corrections to $\Sigma$ arise from two sources:
from the thermal scale, $\mc{O}(\lambda^2)$ corrections to the parameters of the EFT,
and, from the nucleation scale, two-loop corrections in the expansion around the critical bubble.
Regarding the latter, note that the split between dynamical and statistical parts in Eq.~\eqref{eq:divisionDynamicalStatistical} has only been demonstrated to one-loop order within the nucleation scale EFT.

The result, Eq.~\eqref{eq:result_1}, demonstrates the split between the modes above and below the factorisation scale, discussed around Eq.~\eqref{eq:SigmaModes}.
The determinant of fluctuations runs over only those modes that enter the EFT of the nucleation scale (the numerical result can be found in Ref.~\cite{Ekstedt:2021kyx}).
The modes of the thermal scale instead enter the final result through their contribution to the effective parameters of the EFT.
This significantly simplifies the most difficult part of the calculation of $\Sigma$.
In particular, the determinant is evaluated in $\mbb{R}^3$ and not $\mbb{R}^3\times S^1$ and there is no spatially-dependent determinant for the fermion.
In essence, the modes of the thermal scale see the critical bubble as locally constant, with corrections to this picture being suppressed by powers of $\LamNuc/\LamTh$.
Such corrections are accounted for order-by-order by the construction of the EFT in a derivative expansion.


\section{Example 2: Two scale hierarchies} \label{sec:two_scalar}

In this section, we consider the renormalisable model of two $\mathbb{Z}_2$-symmetric scalar fields.
In part of its parameter space, this model realises a first-order phase transition.
However, unlike the previous example, this phase transition is not simply induced by the thermal scale: the first-order nature of the transition is radiatively induced by an intermediate scale.
The same model was analysed in the context of vacuum decay in Ref.~\cite{Weinberg:1992ds}, and in the context of thermal bubble nucleation in Ref.~\cite{Hirvonen:2020jud}, where more detail on the calculations can be found.

The model is defined by the (Euclidean) Lagrangian
\begin{align}
    \mathscr{L} &=\mathscr{L}_\Phi+\mathscr{L}_\Chi+\mathscr{L}_{\rmii{I}} \,,\\
    \mathscr{L}_\Phi&=\frac{1}{2}\left(\partial_\mu \Phi\right)^2+\frac{m^2+\cou{m^2}}{2}\Phi^2+\frac{\lambda+\cou{\lambda}}{4!}\Phi^4 \,,\\
    \mathscr{L}_\Chi&=\frac{1}{2}\left(\partial_\mu \Chi\right)^2+\frac{M^2}{2}\Chi^2+\frac{f}{4!}\Chi^4 \,,\\
    \mathscr{L}_{\rmii{I}}&=\frac{g^2+\cou{g^2}}{4}\Phi^2\Chi^2 \,.
\end{align}
The field $\Phi$ is the nucleating field, and $\Chi$ is the inducing field.
Their three-dimensional counterparts will be denoted $\phi$ and $\chi$ respectively.
In the absence of $\Chi$, the model reduces to a single real $Z_2$-symmetric scalar theory, for which there is only a second-order phase transition.
We have marked only the counterterms that will be necessary up to the order of our calculation.
We will be working in the \MSbar--scheme.

Our first goal is to hone in on the region of the parameter space that contains first-order symmetry breaking transitions of the $\Phi$-field.
To do so it is useful to adopt a formal power counting scheme, in which the magnitudes of quantities are measured in powers of the perturbative coupling $g$ and the temperature $T$.

For the $\Chi$-field to induce a first-order phase transition for $\Phi$, its leading (one-loop) effects must contend with the tree-level potential for the $\Phi$-field.
As a consequence, $\lambda\ll g^2$ is a requirement for a first-order symmetry breaking transition.
However, if we let $\lambda$ be as small as $\lambda \sim g^4$ then the transition becomes so strong as to invalidate the high-temperature expansion.%
\footnote{
    In this case one obtains $\phi_0\sim \sqrt{T}/g$ near the critical temperature and, in the broken phase, the field $\chi$ can receive a correction to its effective mass of order $T$. Relatedly, external $\phi^2$-legs would need to be resummed in the dimensional reduction.
}
Therefore, for simplicity, we will choose from $g^4 \ll \lambda \ll g^2$ the middle road
\begin{equation} \label{eq:lamg3}
\lambda \sim g^3 \,.
\end{equation}
This choice has been motivated along similar lines in studies of first-order phase transitions in gauge-Higgs theories~\cite{Arnold:1992rz,Ekstedt:2020abj}.
In this analogy the $\Chi$-field plays the same role as the gauge fields.

Regarding the remaining parameters of the Lagrangian, we will assume that there are no further anomalously small couplings, so that $f\sim g^2$, and that the high-temperature approximation applies to both particles.
This latter assumption implies that thermal mass corrections are at least as large as the tree-level parameters, and hence that both $m^2$ and $M^2$ are $\order{g^2 T^2}$ or smaller.
For the $\Phi$-field to go through a thermal symmetry breaking transition, its mass parameter should be negative, $m^2<0$.
Conversely, the $\Chi$-field mass parameter should not be too negative, so that it does not go through symmetry breaking before $\Phi$.

From these choices we can determine the orders of magnitude of the necessary quantities related to the critical bubble, following the discussion in Sec.~\ref{sec:formalism}.
Noting that the leading thermal mass correction is $\order{g^2T^2}$, the transition occurs roughly at
\begin{equation} \label{eq:approx_Tc}
    -m^2\sim g^2T^2 \,,
\end{equation}
in the vicinity of which the thermal mass of the $\phi$-field goes through zero.
At such temperatures, we can determine the mass of the nucleating DoF $m_{\text{nucl}}$ and the value of the field at the centre of the bubble $\phi_0$, by demanding that the leading terms in the effective potential of the $\phi$-field are all the same order of magnitude
\begin{equation} \label{eq:ordersOfMagnitudeExampleII}
    m_{\text{nucl}}^2\phi_0^2\sim g^{3}T^{3/2}\phi_0^3\sim\lambda T\phi_0^4 \,.
\end{equation}
Here the cubic term arises from one-loop $\chi$ fluctuations.
Together with Eq.~\eqref{eq:lamg3}, the conclusion is that
\begin{equation}\label{eq:orderTildeMPhi}
   m_{\text{nucl}}^2\sim g^3T^2, \qquad \phi_0\sim T^{1/2} \,.
\end{equation}
The result for $\phi_0$ will be important for noticing the need for resummations in the external $\Phi/\phi$-legs: no need in the dimensional reduction but needed for the second matching step.
While these estimates are valid near the critical temperature, for strongly supercooled transitions nearing spinodal decomposition both $m_{\text{nucl}}^2$ and $\phi_0$ may be smaller than in Eq.~\eqref{eq:orderTildeMPhi}.

Note that $m_{\text{nucl}}^2$ is parametrically smaller than both its constituent parts $m^2$ and $g^2T^2$, due to the approximate cancellation of Eq.~\eqref{eq:approx_Tc}.
No such cancellations occur for the $\chi$-field, and hence
\begin{equation} \label{eq:orderchiMass}
    M_3^2 \sim g^2 T^2 \,.
\end{equation}

The bubble volume is $m_{\text{nucl}}^{-3}\sim g^{-9/2}T^{-3}$, and consequently the nucleation action is $m_{\text{nucl}}^{-1}\phi_0^2\sim g^{-3/2}$ at LO, and is $\sim g^{-1/2}$ at NLO.
To describe the nucleation action up to NLO, we therefore need those contributions to the nucleation scale effective Lagrangian up to and including $\mc{O}(g^4 T^3)$.
This accounts or all the exponentially large contributions to the nucleation rate.
We will also discuss extending the calculation to $\mc{O}(g^5 T^3)$ accuracy, which can become relevant after solving for the dynamical prefactor.

In this model, all three scales of Fig.~\ref{fig:scales} are present: the thermal scale ($\pi^2T^2$), the intermediate scale ($g^2T^2$) and the nucleation scale ($g^3T^2/\pi$).
Here we have reinserted the one-loop factors of $\pi$ in Eqs.~\eqref{eq:lamg3} and \eqref{eq:ordersOfMagnitudeExampleII} to obtain these scales.
The first task is to integrate out the thermal scale, by performing dimensional reduction and thereby creating the EFT for the intermediate scale.
Second, we will need to integrate out the intermediate scale to obtain the nucleation scale EFT.

\subsection{Intermediate scale}

Here we detail the matching relations for dimensional reduction.
The matching relation for the mass parameter of the $\phi$-field and its counterterm reads
\begin{widetext}
\begin{align}\label{eq:phimassmatchex2}
-m_3^2(\Lambda)-\cou{m_3^2} &=
\; \feynalign{\selfenergytree{plain}{crossed dot}} \;
+ \;  \feynalign{\selfenergyoneloop{plain}{dashed}{}} \;
+ \;  \feynalign{\selfenergyoneloop{plain}{plain}{}} \;
+ 
\;  \feynalign{\selfenergyoneloopmassinsert{plain}{dashed}{crossed dot}} \;
+ \;  \feynalign{\selfenergysnowman{plain}} \;
+ \;  \feynalign{\selfenergysnowman{dashed}} \;
+ 
\;  \feynalign{\sunsettwo{plain}{dashed}} \;
+
\; \feynalign{\selfenergytree{plain}{dot}} \;
+ \;  \feynalign{\selfenergyoneloop{plain}{dashed}{dot}{}} \;
+ \;  \feynalign{\selfenergyoneloop{plain}{plain}{dot}{}}
\\
= -m^2-\frac{g^2+\lambda}{2}&\sumint{P}\frac{1}{P^2}
+\frac{g^2}{2}\sumint{P}\frac{1}{P^4}\left(M^2+\frac{g^2+\lambda}{2}\sumint{Q}\frac{1}{Q^2}\right)
+\frac{g^4}{2}\sumint{PQ}\frac{1}{P^2Q^2(P-Q)^2}
-\cou{m^2}-\frac{\cou{g^2}+\cou{\lambda}}{2}\sumint{P}\frac{1}{P^2} \,,
\end{align}
\end{widetext}
where the crossed dot represents a mass insertion and a plain dot represents a counterterm. Note that the momentum dependence of the sunset diagram is already higher order than our desired accuracy.

For the $\chi$-mass, less accuracy is required, as it affects nucleation only via interactions:
\begin{align}
-M_3^2 =& 
\; \feynalign{\selfenergytree{dashed}{crossed dot}} \;
+ \;  \feynalign{\selfenergyoneloop{dashed}{dashed}{}} \;
+ \;  \feynalign{\selfenergyoneloop{dashed}{plain}{}} \;
\\
=& -M^2-\frac{f+g^2}{2}\sumint{P}\frac{1}{P^2} \,.
\end{align}
For the same reason, the self-interaction of the $\phi$-field is the only coupling constant requiring thermal corrections:
\begin{align} \label{eq:lam3_chi_phi_matching}
-\lambda_3/T =& 
\; \feynalign{\fourpointtree{}} \;
+ \; \feynalign{\fourpointcorrection} \;
+ \; \feynalign{\fourpointtree{dot}} \;
\\
=&-\lambda+\frac{3g^4}{2}\sumint{P}\frac{1}{P^4}-\cou{\lambda} \,.
\end{align}

Higher-dimensional operators, resulting from e.g.\ matching 6-point functions, are higher order than the desired accuracy, and so too are the gradient terms resulting from momentum dependent diagrams.

The full result for the dimensionally-reduced effective Lagrangian for the intermediate scale is given by
\begin{align} \label{eq:lagrangian_DR}
    \mathscr{L}_{\rmii{DR}}&=\mathscr{L}_\phi+\mathscr{L}_\chi+\mathscr{L}_{\rmii{I,\,DR}} \,,\\
    \mathscr{L}_\phi&=\frac{1}{2}\left(\grad \phi\right)^2+\frac{m^2(\Lambda)+\cou{m^2_3}}{2}\phi^2+\frac{\lambda_3}{4!}\phi^4 \,,\\
    \mathscr{L}_\chi&=\frac{1}{2}\left(\grad \chi\right)^2+\frac{M_3^2}{2}\chi^2+\frac{f_3}{4!}\chi^4 \,,\\
    \mathscr{L}_{\rmii{I,\,DR}}&=\frac{g_3^2}{4}\phi^2\chi^2 \,,
\end{align}
where the effective fields are identified with the zero Matsubara modes
\begin{align}
    \phi&=\frac{1}{\sqrt{T}}\Phi_0 \,,\label{eq:DRphi2}\\
    \chi&=\frac{1}{\sqrt{T}}\Chi_0 \,,
\end{align}
and the parameters read
\begin{align}
    m_3^2(\Lambda)=&m^2(\mu)+\frac{g^2(\mu)T^2}{24}+\frac{\lambda(\mu) T^2}{24} \nonumber\\
	&-\frac{(g^4+fg^2)T^2+24g^2M^2}{24(4\pi)^2}\ln(\frac{e^{\gE}\bmu}{4\pi T}) \nonumber\\
	&+\frac{g^4T^2}{4(4\pi)^2}\ln(\frac{A^{12}\blam^2}{4\pi T\bmu}) \,,\label{eq:intermediatemassex2}\\
	\cou{m_3^2}=&\frac{g^4T^2}{8(4\pi)^2}\frac{1}{\epsilon} \,,\\
	M_3^2=&M^2+\frac{(g^2+f)T^2}{24} \,,\label{eq:chiMass}\\
	\lambda_3=&\lambda(\mu) T-\frac{3g^4T}{(4\pi)^2}\ln(\frac{e^{\gE}\bmu}{4\pi T}) \,,\label{eq:threeSelfCoupling}\\
	g_3^2=&g^2T \,,\label{eq:threeMixingCoupling}\\
	f_3=&fT \,,
\end{align}
where $A$ is the Glaisher–Kinkelin constant.
Notice, that we have used the one-loop running of the renormalised parameters to eliminate most of the dependence on the factorisation scale $\Lambda$, replacing it with the renormalisation scale $\mu$; see Refs.~\cite{Farakos:1994kx,Braaten:1995cm,Kajantie:1995dw}.
However, there still remains some dependence in $m_3^2(\Lambda)$.
This is important, as it will cancel against the $\Lambda$ dependence of the sunset diagram in the matching for the nucleation scale EFT (and so too will the counterterm).

Physically, the most significant contributions are the leading thermal corrections to the masses, which keep the fields stable in the symmetric phase at high temperatures.
These terms eventually contribute to the nucleation rate at its leading order $\sim\exp(1/g^{3/2})$.
The terms are, in fact, separately of order $\sim\exp(1/g^{5/2})$, but this leading power cancels due to Eqs.~\eqref{eq:approx_Tc} and \eqref{eq:orderTildeMPhi}.
The rest of the terms shown, though smaller, are nevertheless exponentially important contributions to the nucleation rate, contributing at order $\sim\exp(1/\sqrt{g})$.

\subsection{Nucleation scale}

As one varies the temperature, $T$, Eq.~\eqref{eq:lagrangian_DR} does not yet manifestly display a first-order transition -- just a second-order transition. The first-order nature of the transition is radiatively induced by the $\chi$-field. Hence, it will only be visible after integrating out the intermediate scale.

The second step, matching for the nucleation scale EFT, exhibits two distinctions in comparison to the first step of dimensional reduction: It is necessary to resum the external $\phi$-legs, and to include the leading one-loop gradient corrections.

The need for the resummation arises from the fact that a $g^2\phi^2$ insertion to a $\chi$-propagator modifies the order of a diagram by a factor of $g^2\phi^2/M_3^2$. From Eqs.~\eqref{eq:orderTildeMPhi} and \eqref{eq:orderchiMass}, we can see that this is of order one. Hence, the modified diagram is not suppressed compared to the original one. This results in the need to resum external $\phi$-legs in $\chi$-propagators.

Another, equivalent but perhaps more intuitive, viewpoint is that the bubble background gives a LO correction to the $\chi$ mass.
Hence, the effect on the mass has to be taken into account when integrating out the intermediate scale.

The need for gradient corrections arises from the fact that the intermediate scale and the nucleation scale are separated only by one power of $g$.
The one-loop term from the $\chi$-field is of leading order ($g^3T^3$), as it induces the broken minimum for the nucleation scale effective potential.
Thus, the first gradient term is $\mc{O}(g^4T^3)$, which is still exponentially important.

To handle the resummations of the $\chi$-propagator, we reorganise the calculation as a matching of 1$\phi$I-actions, i.e.\ of the generating functionals of all 1$\phi$I $n$-point correlation functions.
Due to the resummations in external $\phi^2$-legs, a brute-force matching of $n$-point functions would be a cumbersome bookkeeping task.

The matching has a particularly simple diagrammatic representation:
\begin{align}
\sn[\phi] =& 
S_{\rmii{DR}}[\phi,\,0]
+ \; \feynalign{\oneloopbubble} \;
+ \; \feynalign{\twoloopbubble} \;
+ \; \feynalign{\sunsetbubble} \,.
\end{align}
The diagrams on the right hand side, in the full theory, are understood to have resummed $\chi$-propagators, and the first gradient correction of the one-loop diagram is included.
The $\phi$-propagator of the sunset diagram is massless, due to the use of strict perturbation theory in $m_3^2$, and it is IR-regulated with dimensional regularisation.

The resulting effective action for the nucleation scale is given by
\begin{align}\label{eq:exampleTwoNuclAction}
	\sn=\int_{\mathbf{x}}&\Biggl\{\frac{Z(\phi)}{2}\qty(\grad{\phi})^2 +\frac{\tilde{m}^2(\phi)}{2}\phi^2\nonumber\\
-\frac{1}{3(4\pi)}&\qty[\qty(M_3^2+\frac{g_3^2}{2}\phi^2)^{3/2}-M_3^3] +\frac{\lambda_3}{4!}\phi^4\Biggr\} \,,
\end{align}
where the parameters are
\begin{align}
	Z(\phi)=&1+\frac{1}{48(4\pi)}\frac{g_3^4\phi^2}{\qty(M_3^2+\tfrac{g_3^2}{2}\phi^2)^{3/2}} \,,\label{eq:WaveFuncRenorm}\\
	\tilde{m}^2(\phi)=&m_3^2(\Lambda)+\frac{f_3g_3^2}{8(4\pi)^2}\nonumber\\
	&-\frac{g_3^4}{4(4\pi)^2}\qty[1-\ln(\frac{4}{\blam^2}\qty(M_3^2+\frac{g_3^2}{2}\phi^2))] \,, \label{eq:exampleTwoNuclMass}
\end{align}
in terms of those in Eqs.~\eqref{eq:chiMass} to \eqref{eq:threeMixingCoupling}.
By inserting the expression for $m_3^2$, we find that all the dependence on the cut-off $\Lambda$ cancels.
So too does the dependence on $\mu$, up to higher order corrections.
The $\Lambda$ dependence and the mass counterterm were cancelled by the sunset diagram.

Here, the field-independent free-energy contribution from the intermediate scale, $M_3^3/12\pi$ (\textit{cf.}\ Eq.\eqref{eq:DRPartFunc}), is explicitly present in the effective action for the nucleation scale.
Due to the one-loop resummation, it cancels only asymptotically on the tail of the critical bubble, setting the action to zero in the metastable phase.

The most crucial contribution from the intermediate scale is the one-$\chi$-loop, the term with square brackets in Eq.~\eqref{eq:exampleTwoNuclAction}.
Together with the tree-level terms, it is responsible for the coexistence of phases, and consequently for the first-order nature of the phase transition.
It thus enters the nucleation rate at its leading order $\sim\exp(1/g^{3/2})$.
The rest of the terms are subdominant, but still exponentially important contributions to the nucleation rate, contributing at $\sim\exp(1/\sqrt{g})$.

Now that we have the effective action for the nucleation scale, Eq.~\eqref{eq:exampleTwoNuclAction}, the general formalism of Sec.~\ref{sec:background} can be put into action to construct $\Sigma$, the statistical part of the nucleation rate.
First one finds the critical bubble configuration, and second evaluates the effective action and the fluctuation determinants about this bubble.

The critical bubble is, as usual, a stationary point of the nucleation scale effective action, Eq.~\eqref{eq:exampleTwoNuclAction}.
Due to the spherical symmetry of the critical bubble~\cite{Blum:2016ipp}, the differential equation simplifies into a one-dimensional differential equation:%
\footnote{
    The equation looks rather unpleasant, but it can be solved numerically, for example, via a generalised version of the method first presented in Ref.~\cite{Espinosa:2018hue}. The straightforward generalisation to field dependent $Z(\phi)$ is given in Ref.~\cite{Hirvonen:2020jud}.
}
\begin{equation}\label{eq:EoMExII}
	Z(\phi)\pdv[2]{\phi}{r}+\frac{2}{r}Z(\phi)\pdv{\phi}{r}+\frac{1}{2}Z'(\phi)\qty(\pdv{\phi}{r})^2 = V_{\rmi{nucl}}'(\phi) \,,
\end{equation}
where $V_{\rmi{nucl}}(\phi)$ is the potential part of the nucleation scale effective action in Eq.~\eqref{eq:exampleTwoNuclAction}.

The above equation of motion contains the NLO contributions from the $\chi$ field, and thereby yields both the critical bubble and the nucleation action up to NLO.
To calculate just the nucleation action up to NLO, one can instead drop these NLO terms from the equation of motion, and merely evaluate them on the critical bubble solved for at LO.
Doing so neglects tadpole corrections, as discussed around Eq.~\eqref{eq:actionleftoutshape}.
These tadpoles contribute to the nucleation action at NNLO order, and hence can be safely neglected, as long as the NLO terms are indeed subleading.
However, this assumption breaks down when the transition has supercooled to $m_{\text{nucl}}^2\sim g^4T^2$, as the NLO contributions to the effective mass then become of LO.
It can also break down in the thin-wall regime, where the NLO contributions to the potential difference between the phases can become the same size as the LO terms.
In both these cases, one must solve the full Eq.~\eqref{eq:EoMExII} to obtain the LO critical bubble.

The result for the statistical part of the nucleation rate can now be read from Eqs.~\eqref{eq:SaddlepointAppr} and \eqref{eq:ZeroModes}:
\begin{align} 
    \Sigma =& 
    V\left[\int_{\mathbf{x}}\frac{(\nabla\cb)^2}{3}\right]^{3/2}\sqrt{\abs{\frac{\det
    (\sn''[\phi_\rmi{meta}]/2\pi)}{\det^\prime (\sn''[\cb]/2\pi)}}}\nonumber\\
    &\times e^{-\sn[\cb]}\left(1 + \mc{O}(\sqrt{g})\right) \,.
    \label{eq:SigmaExample2}
\end{align}
Note that the fluctuation determinants neither bear reference to the compact Euclidean dimension, nor to the field $\chi$, both having been integrated out beforehand.

Parametrically, this result is of the form
\begin{align} 
    \Sigma &= 
    a g^{9/4} T^3 V \exp\left(- \frac{b}{g^{3/2}} + \frac{c}{\sqrt{g}}\right)
    \left(1 + \mc{O}(\sqrt{g})\right) \,,
    \label{eq:parametric_result_2}
\end{align}
where $a$, $b$ and $c$ are $\order{1}$ temperature-dependent values, calculable using the approach presented, and we have used that the low-lying eigenvalues of $\sn''$ scale as $m_{\text{nucl}}^2$.
At LO, and when written in terms of $\lambda$, this parametric result agrees with Eq.~\eqref{eq:parametric_result_1} of Example 1, however there are additional subleading terms in the present example.

The $\order{\sqrt{g}\,\Sigma}$ corrections to $\Sigma$ in Eq.~\eqref{eq:SigmaExample2} come from the missing $\mc{O}(g^5T^3)$ contributions to the effective Lagrangian for the nucleation scale, arising from the thermal and intermediate scales.
The calculation of these becomes relevant if the dynamical part of the nucleation rate is solved to the same accuracy.
Two-loop corrections from the nucleation scale itself are higher order still, contributing to $\Sigma$ at $\order{g^{3/2}\,\Sigma}$.

In the dimensional reduction at $\mc{O}(g^5T^3)$ accuracy there is nothing new computationally:
No new topologies arise, and external momentum dependence is still unimportant.
There are two more diagrams to be included into the $\phi$-mass matching:
\begin{align}
\; \feynalign{\selfenergyoneloopmassinsert{plain}{plain}{crossed dot}} ,\quad \;  \feynalign{\selfenergysnowmanmod{plain}{plain}{dashed}} \,.
\end{align}
There are also two additional contributions to the counterterms $\cou{m^2}$ and $\cou{g^2}$ coming from 4-dimensional Euclidean vacuum diagrams:
\begin{align}
    \;  \feynalign{\selfenergyoneloop{plain}{plain}{}} \;,\quad\;  \feynalign{\fourpointcorrectionmod{plain}} \,,
\end{align}
which cancel the divergences of the corresponding diagrams above.

The matching of the $\chi$-mass must be carried out to $\order{g^4T^2}$, and thus it receives the same complexity as the $\phi$-mass.
The matching includes all the diagrams of the types given in Eq.~\eqref{eq:phimassmatchex2}, except
\begin{align}
\;  \feynalign{\selfenergysnowmanmod{dashed}{plain}{plain}} \;\sim g^5 T^2 \,.
\end{align}
Most notably, $M_3^2$ acquires $\Lambda$ dependence through
\begin{align}
    \;  \feynalign{\sunsettwo{dashed}{dashed}} \,, \quad \;  \feynalign{\sunsettwo{dashed}{plain}} \,.
\end{align}
Similarly to $M_3^2(\Lambda)$, $g_3^2$ needs to be matched to order $\order{g^4T}$.
This means including the following diagrams:
\begin{align}
 \; \feynalign{\fourpointcorrectionmod{dashed}} \,,\quad
 \; \feynalign{\fourpointtreemodded{dot}} \, .
\end{align}
This concludes the relevant corrections at $\order{g^5T^3}$ arising from the thermal scale.

For integrating out the intermediate scale, essentially every modification you can make to a diagram brings an additional power of $g$.
Hence, for $\mc{O}(g^5T^3)$ accuracy, we need to compute all vacuum three-loop diagrams with $\phi$-propagators IR-regulated via dimensional regularisation, an additional gradient term for the one- and two-loop diagrams, and also the sunset diagrams with the insertions to the $\phi$-propagator of the $\phi$-mass or an external $\phi$-leg.

A nontrivial check of the computation at this next order is that the three-loop basketball diagrams should cancel the leading $\Lambda$ dependence from $\sn$ due to $M_3^2(\Lambda)$. 
Of course some $\Lambda$ dependence remains, though this is an $\mc{O}(g^6T^3)$ effect, of higher orders still.

The improvement to $\mc{O}(g^5T^2)$ might seem like a daunting task, and it will only become relevant after solving for the dynamical prefactor. However, the reason for discussing it is to show that the method presented here can be systematically improved to higher orders up to, but not including, two-loop order within the nucleation scale EFT.
Here, this amounts to a correction to $\Sigma$ of order $\order{g^{3/2}\Sigma}$.


\section{Example 3: Scale-shifters} \label{sec:scaleshifters}

It is common that the masses of fields shift during a phase transition due to the changes in vacuum expectation values of the nucleating DoFs.
By scale-shifter, we mean a field whose effective mass changes so drastically that it belongs to parametrically different scales on different parts of the nucleating bubbles, being much heavier than the nucleation scale in at least one part.

In this section, we will discuss how to treat scale-shifters within the EFT approach to bubble nucleation.
More specific discussion here is presented from the perspective of perhaps the cosmologically most relevant case: a symmetry breaking phase transition, in which a field is light (i.e.\ nucleation scale or lower) in the symmetric phase and intermediate in the broken phase (\textit{cf.} Fig.~\ref{fig:scales}).
Examples of this case are strong transitions in gauge-Higgs models~\cite{Arnold:1992rz,Ekstedt:2020abj}, where the spatial components of the gauge field can shift in scales,
and in the cubic anisotropy model, which is a cousin to our model in Sec.~\ref{sec:two_scalar}~\cite{Moore:2001vf}.
We will adopt the cubic anisotropy model as an example for this section.

The heart of the matter with scale-shifters is the following: If they belong to a higher scale on some part of a bubble, they can have a great impact on it, significantly affecting the local free energy density.
Consequently, their contributions should be included in the effective description for the transition surface.
However, since they belong to different scales on different parts of the bubble, they need to be handled in different steps of integrating out on these different parts.

In the symmetry breaking cases discussed above, the scale-shifters would need to be integrated out along with the intermediate scale within the main body of the bubble, but left in the nucleation scale EFT on the bubble tail.
As demonstrated in Example 2,
intermediate scale fluctuations can be of great importance for the critical bubble configuration, making the integration over these modes crucial.
However, if one integrates out the whole scale-shifting field, the result is nonlocal in regions where the field is associated with a lower scale.
The gradient expansion works only on the inside of the bubble, but diverges on the tail.

In attempting to apply an EFT approach to this problem,
first we need to understand at what perturbative order the nonlocal IR physics of the bubble tails contributes to the nucleation action.
The terms of lower order than this are directly calculable within a perturbative EFT approach, whereas the nonlocal terms themselves will require further consideration.

Inspiration for how to proceed can be found in the usual perturbative approach to the study of non-Abelian gauge theories at high temperature.
In that case, the spatial gauge bosons are screened only at $\order{g^2T}$, and render nonperturbative the $\order{g^6 T^4}$ contributions to the free-energy density~\cite{Linde:1980ts}.
However, this does not prevent the perturbative calculation of all the lower order terms, up to $\order{g^6\log{g}\ T^4}$~\cite{Kajantie:2002wa}.
This can be achieved by treating the mass of the spatial gauge bosons in strict perturbation theory, i.e.\ equated to zero within loop integrals.
Doing so projects out the unknown nonperturbative IR physics, while introducing an error of $\order{g^6 T^4}$, which is anyway beyond the reach of perturbation theory.
If one proceeds further within strict perturbation theory, one finds that the $\order{g^6 T^4}$ and higher order terms are IR divergent, though physically these divergences are screened by the mass of the spatial gauge bosons.
This analogy sheds light on much of the following.

In the present case, an analogous procedure would be to treat the light IR quantities of the symmetric phase strictly perturbatively, when integrating out the scale-shifter.
This projects out the difficult nonlocal physics from the lowest few orders, at the cost of introducing errors at higher orders.
It is made possible because the light physics near the symmetric phase has only a higher-order effect on the nucleating DoFs, and the IR quantities can be treated perturbatively on the body of the bubble due to the scale hierarchy, $\Lambda_{\rmii{light}}\ll\Lambda_{\rmii{int}}$.

In the following, we will test the validity of the strict perturbative expansion as applied to scale-shifters in the cubic anisotropy model.
We show that it can be linked to the perturbativity of the nucleating DoF in the symmetric phase.
We will see that at LO this approach is straightforward, and yields the parametrically largest contributions to the nucleation scale effective theory.
Further, NLO corrections in this expansion yield the dominant subleading contributions.
However, while they are finite and computable, the NLO corrections from the scale-shifters will begin to reveal the IR complications of the symmetric phase.
They should only be evaluated on the leading order critical bubble, dropping higher order corrections coming from their effect on the critical bubble.
Beyond NLO there are IR divergences in strict perturbation theory, though physically these divergences are screened.
The irreducible error in this expansion is due to the contributions of nonlocal infrared physics, which can be estimated parametrically.

The strict perturbative expansion projects out all the IR physics of the symmetric phase, and not just the nonlocal IR physics.
In particular, it misses the normalising partition function evaluated around the metastable (symmetric) phase in Eq.~\eqref{eq:GeneralArea}.
In our Example 2 model, this gave the mass-cubed term $M_3^3/12\pi$ in Eq.~\eqref{eq:exampleTwoNuclAction}.
Note, that its effect is to make the nucleation scale effective action
asymptote to zero in the symmetric phase.
Due to the strict expansion, the term is set identically to zero.
The error yielded by the absence of this term is noteworthy only when the scale-shifter mass is of the nucleation scale (and not lighter), and in these cases it may actually give the largest error of the strict perturbation expansion.

At the end of this section, we will discuss ways to go beyond the strict perturbative expansion.
Most notably, we will show that the full error-free exponential order is obtainable with the help of numerical methods for evaluating the scale shifter one-loop term on the LO critical bubble.
Also, we will show how to eradicate the error coming from the missing mass-cubed term via a resummation, and other tricks to soften the singular behaviour in derivatives of the action obtained through the strict expansion.

Before moving further, let us briefly overview the cubic anisotropy model, to give
a concrete example containing a scale-shifter.
The model is akin to our Example 2, except that the scalars, $\Phi$ and $\Chi$, are symmetric, $M^2=m^2<0$ and $f=\lambda\ll g^2$.%
\footnote{
    The couplings are chosen so that the broken phases exist in the field directions corresponding to the field variables.
    There is clearly freedom to e.g.\ rotate the field basis by e.g.\ $\pi/4$, moving the locations of the broken minima (see for example Ref.~\cite{Moore:2001vf}).
}
Hence, both of the fields can nucleate during the phase transition.
The symmetry between the fields is broken by a nucleating bubble.
We will use the freedom to label the nucleating DoF forming the bubble as $\phi$, and consequently the inducing DoF as $\chi$, matching the convention of Sec.~\ref{sec:two_scalar}.
The power counting turns out exactly the same as that in Sec.~\ref{sec:two_scalar}, but with $f\sim g^3$, and the $\chi$ mass in the symmetric phase being the same as the $\phi$ mass.

We will continue to denote the symmetric phase $\chi$ mass as $M_3$, even though there is a symmetry between the $\phi$ and the $\chi$ masses: $M_3^2=m_3^2$.
(Note that only the leading $g^3T^2$ accuracy is needed for $M_3^2$, but $g^4T^2$ accuracy is needed for $m_3^2$, to achieve the NLO accuracy of this section.)
This will make it easier to see what is specific to the cubic anisotropy model, and what will differ in other models.
For example, in non-Abelian Higgs models, the spatial components of the gauge field play the role of the scale-shifter, and their squared mass in the symmetric phase is lighter, being of $\order{g^4T^2}$.
While the calculations presented below will be specific to the cubic anisotropy model, the same arguments and methods will apply also to gauge-Higgs models.

\subsection{Leading order critical bubble}

Here, we will utilise the strict perturbative expansion to find the LO critical bubble in the cubic anisotropy model.
The local LO action will inevitably contain errors near the symmetric phase.
These will be shown to be negligible as long as the nucleating DoF remains perturbative in the symmetric phase.

From our Example 2, we can read off the LO contribution to be
\begin{align}\label{eq:scaleShiftersStrictOneLoop}
\;  \feynalign{\oneloopbubble} \; = 
\int_{\mathbf{x}}\frac{-g_3^3}{6\sqrt{2}(4\pi)}|\phi|^3
\,.
\end{align}
This is the same as the one-loop potential contribution in Example 2, except with just the scale-shifting part of the $\chi$ mass, $M_3^2(\phi)=g_3^2\phi^2/2$, i.e.\ without the constant additive contribution, $M_3^2$.
The full LO action is thus%
\footnote{
    Note that, to avoid notational clutter, the split presented here is $\Lambda$ dependent. Moving the $\order{g^4}$ part of $m_3^2(\Lambda)$ to the NLO action shown in Eq.~\eqref{eq:scaleShiftersNLOAction} would make the split cut-off independent.
    The $\epsilon$-pole has also been implicitly cancelled.
}
\begin{align}\label{eq:scaleShiftersLOAction}
	\sn^{\rmii{(LO)}} = \int_{\mathbf{x}}&\biggl\{
	\frac{1}{2}\qty(\grad{\phi})^2
	+\frac{m_3^2(\Lambda)}{2}\phi^2 
	 \nonumber \\
    &\qquad -\frac{g_3^3}{6\sqrt{2}(4\pi)}|\phi|^3
    +\frac{\lambda_3}{4!}\phi^4\biggr\}
    \,,
\end{align}
where the masses and coupling constants can be found from Sec.~\ref{sec:two_scalar} in Eqs.~\eqref{eq:intermediatemassex2}, \eqref{eq:chiMass}--\eqref{eq:threeMixingCoupling} (though $f\sim g^3$ renders some terms negligible), and the equation of motion for the critical bubble is given by
\begin{equation}\label{eq:ScaleShifterEoM}
	\pdv[2]{\phi}{r}+\frac{2}{r}\pdv{\phi}{r} = V^{\rmii{(LO)}\,\prime}_{\rmi{nucl}}(\phi)
	\,,
\end{equation}
where $V^{\rmii{(LO)}}_{\rmi{nucl}}(\phi)$ is the potential part of the LO nucleation scale effective action in Eq.~\eqref{eq:scaleShiftersLOAction}.
When evaluated on the critical bubble, the LO nucleation scale effective action is of order $1/g^{3/2}$.

Underlying Eqs.~\eqref{eq:scaleShiftersStrictOneLoop} to \eqref{eq:ScaleShifterEoM}, and the strict perturbation expansion, is the assumption that the $\chi$ field is of an intermediate scale, $M_3(\phi)\gg m_{\text{nucl}}$, yet this is only true for sufficiently large values of $\phi$.
To estimate the errors introduced by this assumption, it is therefore necessary to discover how small values of $\phi$ can be reached, while $\chi$ remains of the intermediate scale.
We will see that the nonlocality only begins on the bubble tail, where the $\tfrac{1}{2}m_3^2\phi^2$ term dominates over the $\chi$ contributions, as long as the nucleating DoF, $\phi$, is perturbative in the symmetric phase.
Whereupon the nonlocal errors are under control, as a negligible contribution to Eqs.~\eqref{eq:scaleShiftersStrictOneLoop}--\eqref{eq:ScaleShifterEoM}.

To investigate this in detail, it will be useful to first consider the region near the top of the potential barrier, where the $\phi$-derivative of the mass term and the LO one-loop term are exactly opposite, because there the two terms are equal in magnitude. Here, the $\chi$ field contributions are not subdominant, and the field has to have already shifted to the intermediate scale, $M_3(\phi_{\rmi{top}})\gg m_{\text{nucl}}$, for the description of the transition surface to be valid.
Denoting this point by $\phi_{\rmi{top}}$, one arrives at the following equality
\begin{align}
    m_{\text{nucl}}^2\phi_{\rmi{top}}&=\frac{g_3^2}{8\pi}M_3(\phi_{\rmi{top}})\,\phi_{\rmi{top}} \,,\\
    \Rightarrow
    M_3(\phi_{\rmi{top}})&=\frac{8\pi m_{\text{nucl}}}{g_3^2}\cdot m_{\text{nucl}} \,.
    \label{eq:M3Top}
\end{align}
Eq.~\eqref{eq:M3Top} relates the masses of the $\chi$ and $\phi$ fields near the top of the potential barrier.

In the approach towards the symmetric phase, for sufficiently small $\phi\lesssim\phiNonlocal$ the masses of $\chi$ and $\phi$ will be of the same order, and hence the result of integrating out $\chi$ will be nonlocal.
This occurs at
\begin{align}
    M_3(\phiNonlocal) &= m_{\text{nucl}} \,, \\
    \Rightarrow
    \phiNonlocal &= \frac{g_3^2}{8\pi m_{\text{nucl}}} \cdot \phi_{\rmi{top}} \,.
    \label{eq:phiNonLocal}
\end{align}
Note, that here the local EFT fails to describe the $\chi$ contributions accurately, thus they must be subdominant for the validity of our approach.
This is satisfied if $\phiNonlocal \ll \phi_{\rmi{top}}$.

Now, perturbativity of the nucleating field in the symmetric phase implies that
\begin{align}\label{eq:scaleshiftersgood}
    \frac{g_3^2}{8\pi m_{\text{nucl}}}\ll 1 \,.
\end{align}
This combination appears in both Eq.~\eqref{eq:M3Top} and Eq.~\eqref{eq:phiNonLocal} above, allowing us to reach the following conclusions, when Eq.~\eqref{eq:scaleshiftersgood} holds:
the $\chi$ field is of an intermediate scale, when the $\chi$ contributions are not subdominant, $\phi\gtrsim\phi_{\rmi{top}}$,
and the $\chi$ contributions are subdominant, when it is associated with the nucleation scale, $\phi\lesssim\phiNonlocal$.
Therefore, the strict perturbative expansion is valid at least at leading order, when the nucleating field is perturbative in the symmetric phase.%
\footnote{
    Note, that $g_3^2/(4\pi m_3) \approx 1$ in the benchmark point of Ref.~\cite{Moore:2001vf}, so perturbation theory is not reliable at the phase transition.
}
The magnitude of errors associated with the region $\phi\lesssim\phiNonlocal$ is determined in Sec.~\ref{sec:scale_shifter_errors} below.

\subsection{Strict perturbation theory: Next-to-leading order}

Next, we will examine the NLO corrections from the scale-shifter $\chi$ in strict perturbation theory.
These contribute to the nucleation rate at exponential order, $\exp(1/\sqrt{g})$,
and consist of corrections to the one-loop term from both the light-mass expansion (in the cubic anisotropy model) and the gradient expansion, and also
the two-loop sunset correction to the potential.
By the analogous power counting to Example 2, these are the last exponential contributions to the statistical part of the nucleation rate.
Below, we will also find that these are larger than the nonlocal errors from the bubble tail.

The NLO correction from the light-mass expansion is
\begin{equation}\label{eq:MassCorrToCAM}
     \;  \feynalign{\cammassinsert} \; = \int_{\mathbf{x}}\frac{-g_3 M_3^2}{2\sqrt{2}(4\pi)}\abs{\phi} \,,
\end{equation}
where the $\chi$ mass parameter, denoted with $M_3$, is needed only to its leading order, $\order{g^3T^2}$.
Note that this term would yield a singular contribution to the effective mass, $V_{\text{eff}}''(\phi=0)$, and move the physical metastable minimum away from $\phi=0$.
Thus, it cannot be included in the equation of motion for the critical bubble, but can only be evaluated on the LO critical bubble.

The two other NLO contributions are the first gradient correction from the one-loop term,
\begin{equation}\label{eq:ZsingCAM}
    \; \feynalign{\oneloopbubble}_{\text{1-grad}} \; = \int_{\mathbf{x}}\frac{\sqrt{2}}{48(4\pi)}\frac{g_3}{\abs{\phi}}(\grad \phi)^2 \,,
\end{equation}
and the two-loop sunset,
\begin{align}\label{eq:logsinginCAM}
\; \feynalign{\sunsetbubble} \;= \int_{\mathbf{x}} -\frac{g_3^4}{4(4\pi)^2}\qty[\frac{1}{4\epsilon}+\frac{1}{2}-\ln(\frac{\sqrt{2}g_3\abs{\phi}}{\blam})]\phi^2 \,.
\end{align}
There are also singularities in the second $\phi$-derivatives of these terms, from the field renormalisation and the logarithm respectively.
Note that the $\epsilon$ pole and the cut-off dependence cancel the same way as in Sec.~\ref{sec:two_scalar}, between the sunset diagram and the counterterms and parameters of the effective action for the nucleation scale. The cancelling cut-off dependence is shown explicitly in Eq.~\eqref{eq:scaleShiftersLOAction}.

The NLO corrections are obtained by evaluating the LO critical bubble from Eq.~\eqref{eq:ScaleShifterEoM} on the terms in Eqs.~\eqref{eq:MassCorrToCAM}, \eqref{eq:ZsingCAM}, and \eqref{eq:logsinginCAM},
\begin{align}\label{eq:scaleShiftersNLOAction}
	\sn^{\rmii{(NLO)}} &= \int_{\mathbf{x}}\Biggl\{
	\frac{\sqrt{2}}{48(4\pi)}\frac{g_3}{\abs{\cb^{\rmii{(LO)}}}}(\grad \cb^{\rmii{(LO)}})^2  
	-\frac{g_3 M_3^2}{2\sqrt{2}(4\pi)}\abs{\cb^{\rmii{(LO)}}}
	 \nonumber \\
    &\quad -\frac{g_3^4}{4(4\pi)^2}\qty[\frac{1}{2}-\ln(\frac{\sqrt{2}g_3\abs{\cb^{\rmii{(LO)}}}}{\blam})](\cb^{\rmii{(LO)}}){}^2 \Biggr\} \,.
\end{align}

In this approach, of evaluating NLO terms on the critical bubble rather than including them in the equation of motion,
we have uncancelled tadpole contributions.
In the perturbative expansion, diagrams containing tadpoles relate to
changes of the critical bubble configuration (see for example Ref.~\cite{Fukuda:1975di}).
We can confirm that the shape changes are indeed NNLO by estimating the order of the leading (two-loop) tadpole contribution to the nucleation action.
When $\LamNuc\sim m_{\text{nucl}}\sim g^{3/2}T$, this is
\begin{equation}\label{eq:scaleshiftertadpole}
    \feynalign{\shapechangeScaleShifter} \sim \sqrt{g} \,,
\end{equation}
which is indeed NNLO.
The same power counting holds for the other NLO tadpoles as well.

Lastly, we would like to note that evaluating the fluctuation determinants of the $\phi$ field using the LO action gives the determinants correctly to their leading order, even though their contribution is of slightly higher order than NLO.
This follows simply from the fact that everything that enters the evaluation of the determinants, the nucleation scale action and the critical bubble, are correct to their leading order.

\subsection{Strict perturbation theory: Error analysis} \label{sec:scale_shifter_errors}

Next, we investigate the size of the errors coming from the $\chi$ contributions when the field is associated with the nucleation scale, $\phi\lesssim\phiNonlocal$. First, we will find the error from the LO $\chi$ one-loop term, and then analyse the errors linked to higher-order terms using the strict perturbation and loop expansion parameters.

We will investigate the size of the errors coming from the one-loop term directly by estimating the contribution to Eq.~\eqref{eq:scaleShiftersStrictOneLoop} from the region where the $\chi$ field is of the nucleation scale,
\begin{equation}\label{eq:cubicerrorestim}
    \int_{\rNonlocal}^\infty\dd r 4\pi r^2 \left(\frac{-g_3^3}{6\sqrt{2}(4\pi)}\right)|\phi|^3 \,,
\end{equation}
where $\rNonlocal$ is the radius at which the $\chi$ joins the nucleation scale, and hence beyond which its effects on the nucleation scale are nonlocal. The radius can be estimated with
\begin{equation}\label{eq:joiningnuclscal}
    g_3^2\cb^2(\rNonlocal)\sim m_{\text{nucl}}^2,
\end{equation}
where the left-hand side is the scale-shifting part of the $\chi$ mass, $M_3^2(\phi)$. This is equivalent to $\phi\sim\phiNonlocal$.

When the perturbativity condition in Eq.~\eqref{eq:scaleshiftersgood} holds, the $\phi$ mass term $m_{\text{nucl}}^2\phi^2/2$ dominates in the potential at $\phi\lesssim\phiNonlocal$ and the critical bubble profile is given by
\begin{equation}\label{eq:bubbletailforscsh}
    \cb(r) = \frac{A_\infty}{r}e^{-m_{\text{nucl}}r},
\end{equation}
where $A_\infty$ is a constant.
Using Eq.~\eqref{eq:joiningnuclscal} with Eq.~\eqref{eq:bubbletailforscsh}, we obtain
\begin{equation}\label{eq:estimrtilde}
    \rNonlocal\sim m_{\text{nucl}}^{-1}\log{\frac{1}{g}}\qquad\text{or}\qquad\rNonlocal\sim m_{\text{nucl}}^{-1} \,,
\end{equation}
depending on the parametric form of $A_\infty$, which is beyond the current scope.%
\footnote{
    We want to note here that, in the thin wall regime, $A_\infty \propto e^{m_{\text{nucl}}R}$, thereby introducing the correct, thin-wall regime nucleation scale, $R$, (see Appendix~\ref{appendix:thinwall}) to the estimates of $\rNonlocal$ in Eq.~\eqref{eq:estimrtilde} below.
}

Due to the integral being cut off at large radii by the exponential suppression of the bubble tail, $\exp{-m_{\text{nucl}}r}$, we can estimate
\begin{align}
    r &\sim \rNonlocal \,,\\
    \int_{\rNonlocal}^\infty\dd r&\sim m_{\text{nucl}}^{-1} \,,\\
    \cb &\sim \frac{m_{\text{nucl}}}{g_3} \,,
\end{align}
where the last estimate follows in conjunction with Eq.~\eqref{eq:joiningnuclscal}.

With these estimates, we immediately obtain from Eq.~\eqref{eq:cubicerrorestim} that the contribution to the nucleation action from the region where $\chi$ is light is of order
\begin{equation}\label{eq:cubicerrfound}
    \order{(\rNonlocal\,m_{\text{nucl}})^2} \,,
\end{equation}
where the order of $\rNonlocal$ is given in Eq.~\eqref{eq:estimrtilde}.
Thus it is at least $\order{1}$ and at most $\order{(\log g)^2}$.
This provides an estimate of the error in the LO term \eqref{eq:scaleShiftersStrictOneLoop} due to the nonlocality of $\chi$ on the bubble tail.
In essence, it is a $\sim\mNuc^3/(4\pi)$ error in the Lagrangian density, coming from a spherical shell of thickness $1/m_{\text{nucl}}$, and at a radius $\rNonlocal$ on the bubble tail.

We can use the expansion parameters of the $M_3^2$, gradient and coupling expansion to determine the estimates for the errors of the higher order terms, on the basis of the leading order error given in Eq.~\eqref{eq:cubicerrfound}. The expansion parameters of the two former expansions, i.e.\ the expansion in the light quantities, are given by
\begin{equation}\label{eq:lightExpansionParams}
    \frac{\grad{}^2}{\frac{1}{2}g_3^2\phi^2}, \qquad \frac{M_3^2}{\frac{1}{2}g_3^2\phi^2} \,,
\end{equation}
and for the coupling expansion by
\begin{equation}\label{eq:couplingExpansionParamsScaleSh}
    \frac{g_3^2}{4\pi g_3\phi}, \qquad \frac{f_3}{4\pi g_3\phi} \,,
\end{equation}
where the mass parameter in the denominator comes from the scale-shifting part of the $\chi$ mass, $\frac{1}{2}g_3^2\phi^2$, due to the strict perturbation expansion in $M_3^2$ and the $\phi$ mass, $m_3^2$.

Let us first look at the expansion in the light quantities, Eq.~\eqref{eq:lightExpansionParams}, at one-loop level.
Note that on the main body of the bubble, $\phi\sim\phi_{\rmi{top}}$, where $\chi$ is of the intermediate scale, both expansion parameters are $\mc{O}(g_3^4/(8\pi m_{\text{nucl}})^2)=\mc{O}(g/4\pi)$, so that the expansion in the light quantities is well behaved.
When the $\chi$ field joins the nucleation scale at $r\gtrsim\rNonlocal$, these expansion parameters are $\order{1}$.
For this reason, all the light expansion terms from the one-loop level yield parametrically the same error estimate as the leading term, Eq.~\eqref{eq:cubicerrfound}.

The problem we then face is the question of whether the error from one-loop level is truly under control.
Let us consider first the region where $\phi\sim\phiNonlocal$, and hence the expansion parameters of Eq.~\eqref{eq:lightExpansionParams} are of order one.
In this case, there exists only one scale in the full one-loop diagram -- the nucleation scale.
Simply by dimensional analysis, we can then conclude that the full one-loop diagram is of order $\order{m_{\text{nucl}}^3/(4\pi)}$.
This is the same as our estimate based only on the term Eq.~\eqref{eq:scaleShiftersStrictOneLoop}.
Hence, for $\phi\sim\phiNonlocal$, the full error is of the same order of magnitude as that of all the individual terms in the expansion powers of $M_3^2$ and $\grad{}^2$.
Thus, the previous error estimates are unchanged, and the errors are indeed under control.

Even further from the bubble centre, at $\phi\ll\phiNonlocal$, the expansion parameters of Eq.~\eqref{eq:lightExpansionParams} become much larger than one.
Again, we should look at the full diagram.
Now, the scale-shifting part of the $\chi$ mass is much smaller than the symmetric phase mass, $g_3^2\phi^2/2\ll M_3^2$, and hence its resummation is no longer needed.
The leading order in the $\phi^2$-leg expansion is given by the plain, $\chi$ one-loop diagram around the metastable phase, and it cancels identically against the determinant from $Z_{\rmi{meta}}$.
The next contribution, the $\chi$ one-loop diagram dressed with a $\phi^2$-leg, gives a contribution to the nucleation action which is much smaller than $\order{1}$, and can hence be neglected.

At two-loop level, the leading contribution comes from the sunset diagram in Eq.~\eqref{eq:logsinginCAM}.
We could repeat the same analysis as at one-loop, but it is much more straightforward to deduce the result by considering the coupling expansion parameters of Eq.~\eqref{eq:couplingExpansionParamsScaleSh}.
At $\phi\sim\phiNonlocal$, they are $\sqrt{g}/4\pi$ and $g^{3/2}/4\pi$ respectively.
Thus, the leading error from two-loop level is suppressed by $\sqrt{g}\log g/4\pi$ compared to the one-loop error of Eq.~\eqref{eq:cubicerrfound}.
The argument also extends to higher loop orders.

As already noted above, the expansion parameters of the strict perturbative expansion, Eqs.~\eqref{eq:lightExpansionParams}, \eqref{eq:couplingExpansionParamsScaleSh}, diverge far enough from the centre of the critical bubble.
These infrared divergences are not physical in nature, but are artefacts of the strict perturbative expansion not being applicable.
Furthermore, the LO and NLO contributions to the critical bubble and its effective action come from within a region in which the strict perturbative expansion is well behaved.
Therefore, the IR divergences do not nullify the LO and NLO contributions obtained.

There is another source of error, which was mentioned when initially discussing strict perturbation theory.
It is the cubic mass term, $M_3^3/12\pi$, in the $\chi$ one-loop contribution of Eq.~\eqref{eq:exampleTwoNuclAction}.
This term is projected out by strict perturbation theory, even though it is not actually problematic due to nonlocality: It is just the plain one-loop contribution of $\chi$ to the free energy of the symmetric phase.
Still, a naive reintroduction of the term to the strict perturbative expansion would lead to a divergence with volume, because the effective action would no longer asymptote to zero at spatial infinity.
We will discuss a way to consistently reintroduce this term in Subsec.~\ref{subsec:beyondstrict}.

Let us now parametrically estimate the error from the missing cubic mass term of the $\chi$ field. 
Considering first the region where $\phi\ll\phiNonlocal$, we note that at LO the full one-loop diagram evaluated on the critical bubble cancels identically with the one-loop diagram evaluated in the metastable phase.
This point was in fact already made above.
Thus, the dominant contribution from the cubic mass term arises from the region $\phi\gtrsim\phiNonlocal$.
This error can be estimated as
\begin{equation}\label{eq:cubicMassTermErr}
    \tfrac{4\pi}{3}\rNonlocal^3\times M_3^3/12\pi=\order{(\rNonlocal M_3)^3} \,.
\end{equation}
Note, that this is the leading error for the strict perturbative calculation in the cubic anisotropy model if $\rNonlocal\sim m_{\text{nucl}}^{-1}\log \tfrac{1}{g}$.

Concluding this subsection, we arrive at the following parametric result using strict perturbation theory, for the statistical part of the nucleation rate in the cubic anisotropy model
\begin{align} 
    \Sigma &= 
    a g^{9/4} T^3 V \exp\left(- \frac{b}{g^{3/2}} + \frac{c}{\sqrt{g}} + \order{\rNonlocal^3m_{\text{nucl}}^3}\right) \,.
    \label{eq:parametric_result_strict_pert}
\end{align}
The two leading exponential terms, $b$ and $c$, amount to the LO and NLO terms discussed above.
The result is rather similar to that of Example 2, except that
there are still errors present in the exponent that are at least of $\order{1}$ and at most of $\order{(\log g)^3}$.
Note that rather amusingly the statistical prefactor, $a$, is computable and under control at its leading order, even though it is subdominant to the residual errors in the exponent.

\subsection{Beyond the strict expansion}\label{subsec:beyondstrict}

Now, that we have discussed how to obtain the statistical part of the nucleation rate within the pure strict expansion, we would like to discuss alternative means to push the accuracy even higher.
First, we will show how to handle the error from the cubic mass term, Eq.~\eqref{eq:cubicMassTermErr}, and then, how to eradicate all the errors from one-loop level, leaving only the subexponential error from the nonlocal part of the two-loop sunset diagram.
This is the crowning achievement of the section.
At the end, we also discuss ways to soften the singular behaviour of derivatives of NLO terms.
This can become important in the thin-wall regime.

As already discussed above, the cubic mass term $M_3^3/(12\pi)$ is absent in the strict perturbative expansion, and it cannot naively be added because doing so would mean that the effective action would not asymptote to zero at spatial infinity, leading to a volume divergence.
Consequently, we need to step out of the strict perturbative expansion in $M_3^2$ to obtain the correct asymptotic behaviour when the cubic mass term is included.

The cubic mass term is nonanalytic in $M_3^2$, and hence obtaining it requires going to all orders in the $M_3^2$ expansion, though only to one-loop in the loop expansion.
This calculation was in fact already carried out in our Example 2.
The one-loop term in the square brackets in Eq.~\eqref{eq:exampleTwoNuclAction} is the $M_3^2$-resummed one-loop potential.
It correctly asymptotes to zero at spatial infinity.
Using this resums the one-loop potential in $M_3^2$, which mixes orders in the strict perturbation expansion.
Compared to the strict perturbative treatment, it is
\begin{align}\label{eq:scaleShiftersUnstrictOneLoop}
&\left[\int_{\mathbf{x}}\frac{-g_3^3}{6\sqrt{2}(4\pi)}|\phi|^3\right]_{\text{LO}}+\left[\int_{\mathbf{x}}\frac{-g_3 M_3^2}{2\sqrt{2}(4\pi)}\abs{\phi}\right]_{\text{NLO}}\nonumber\\
\longrightarrow &\int_{\mathbf{x}}-\frac{1}{3(4\pi)}\qty[\qty(M_3^2+\frac{g_3^2}{2}\phi^2)^{3/2}-M_3^3]_{\text{LO}}\,,
\end{align}
where the latter term can be read from Eq~\eqref{eq:exampleTwoNuclAction}.
The subscripts LO and NLO indicate how the terms should be treated, when finding the LO critical bubble.

The modification of Eq.~\eqref{eq:scaleShiftersUnstrictOneLoop} correctly accounts for the $\order{\rNonlocal^3\mNuc^3}$ term discussed around Eq.~\eqref{eq:cubicMassTermErr}, while introducing no new errors at lower parametric order.
Hence the leading error which remains is due to nonlocal physics on the bubble tail, Eq.~\eqref{eq:cubicerrfound}.
The parametric result is now
\begin{align} 
    \Sigma &= 
    a g^{9/4} T^3 V \exp\left(- \frac{b}{g^{3/2}} + \frac{c}{\sqrt{g}} + \order{\rNonlocal^2m_{\text{nucl}}^2}\right) \,. 
    \label{eq:parametric_result_mass_resummed}
\end{align}
Note that this residual error is due purely to $\chi$ fluctuations at one-loop order.

Next, we will discuss how to eradicate this one-loop error altogether.
Note that it is due to the nonlocal physics of the bubble tail, containing derivative corrections from all orders in the expansion in $\grad{}^2$.
Thus, the gradient expansion must be resummed.

A crucial observation for computing the full $\chi$ one-loop term is that the critical bubble obtained with the action in Eq.~\eqref{eq:scaleShiftersLOAction} is the correct critical bubble configuration at LO.
This means that the tadpole contributions from the $\chi$ one-loop diagram cancel up to NNLO, as can be seen from Eq.~\eqref{eq:scaleshiftertadpole}.
With strict perturbation theory, we used this to evaluate the NLO contributions on the critical bubble solution.
In addition, one can use this to evaluate the full one-$\chi$-loop term as functional determinants, without making any derivative or mass expansion.
Numerical methods for doing so are briefly discussed in Appendix~\ref{appendix:numerics}.

Algorithmically, the procedure is as follows.
First, one finds the LO critical bubble using the action in Eq.~\eqref{eq:scaleShiftersLOAction}, or with an improved method e.g.\ using Eq.~\eqref{eq:scaleShiftersUnstrictOneLoop}.
Next, one evaluates the LO nucleation scale effective action without the $\chi$ contributions, as well as the two-loop sunset of Eq.~\eqref{eq:logsinginCAM} on the LO critical bubble.
Finally, one evaluates the $\chi$ one-loop functional determinants, in the background of the LO critical bubble.

In this way, at one-loop order one resums both the $M_3^2$ and the $\grad{}^2$ expansions to all orders.
This correctly handles both the cubic mass term of the symmetric phase and the nonlocalities at one-loop level.
The remaining uncertainty due to the nonlocal part of the two-loop sunset is subexponential.
Consequently, the fluctuation determinants of the $\phi$ field, and the dynamical part of the nucleation rate, become accessible accuracy-wise.
Another way to put it is that the full leading order of the nucleation rate is accessible, once the dynamical part can be solved for. 
The parametric form of the result is then exactly the same as in Example 2,
\begin{align} 
    \Sigma &= 
    a g^{9/4} T^3 V \exp\left(- \frac{b}{g^{3/2}} + \frac{c}{\sqrt{g}}\right)
    \left(1 + \mc{O}(\sqrt{g})\right) \,.
    \label{eq:parametric_result_3}
\end{align}
Eq.~\eqref{eq:parametric_result_3} marks the pinnacle of this section, and demonstrates that the EFT method can be applied equally successfully to scale-shifters as to more straightforward heavy DoFs, such as the $\chi$ field in Example 2.

Lastly, we want to discuss a more minor point: softening the singularities of derivatives of NLO corrections. This has already been achieved for the linear potential term, Eq.~\eqref{eq:MassCorrToCAM}, through the replacement of Eq.~\eqref{eq:scaleShiftersUnstrictOneLoop}, thereby allowing the inclusion of this term into the equation of motion for the critical bubble, i.e.\ into Eq.~\eqref{eq:ScaleShifterEoM}.

While softening singularities in the second derivatives of NLO terms does not generically lead to any parametric improvements in the calculation of the nucleation rate, it can become crucial in the thin-wall regime.
In this case, due to the anomalous smallness of the free energy difference between phases, higher loop order corrections to the potential can be of leading order in magnitude.
Resumming the corresponding tadpole corrections is then essential to obtain the LO critical bubble.

In the cubic anisotropy model, this softening can be naturally achieved by resumming $\chi$ mass contributions, as in Eq.~\eqref{eq:scaleShiftersUnstrictOneLoop}.
This can be carried out for all three NLO Lagrangian terms, and yields the same parametric forms as in Example 2.
This resummation was carried out for the one-loop terms in Ref.~\cite{Moore:2001vf}.
In addition, there a renormalisation group improvement was carried out, which alleviates the singularity in the second derivative of the sunset diagram.
In the $\mathrm{SU}(2)$ gauge-Higgs model, Ref.~\cite{Moore:2000jw} proposed a modification of the field renormalisation in the region $\phi\lesssim \phiNonlocal$, so as to soften its singular behaviour there, without affecting the leading behaviour at larger $\phi$.
In so doing, a formally higher-order error was introduced, and by varying the functional form of the modification, the size of this error could be estimated.


\section{Discussion} \label{sec:discussion}

In this article, we have put forward a general approach to calculate the rate of thermal bubble nucleation from first principles, up to a factorised dynamical prefactor.
The idea essentially fleshes out Langer's blueprint of an equation, Eq.~\eqref{eq:LangerFree},
with the necessary details for application to relativistic QFTs at high-temperature.
As such, the EFT framework provides a compelling physical picture of high-temperature bubble nucleation.

The EFT approach also provides a powerful computational framework.
Crucially, it provides a self-consistent way out of the apparent catch-22 which has plagued calculations of thermal bubble nucleation in the particle physics literature.
In addition, it facilitates the identification of all exponentially large contributions to the nucleation rate, including those arising at two-loop and higher order, and simplifies their calculation, by treating only one scale at a time.

To put the present article in context, in the following we compare the EFT approach to existing approaches in the literature, focusing on their advantages and disadvantages for concrete calculations.
In doing so we provide a brief overview of the literature on the subject.

\approachtitle{Naive approach}
Following Linde's early work~\cite{Linde:1980tt,Linde:1981zj}, by far the most common approach taken in the literature (see for example Refs.~\cite{Delaunay:2007wb,Espinosa:2008kw,Funakubo:2009eg,Wainwright:2011kj,Beniwal:2017eik,Huang:2020bbe,Borah:2020wut,Cline:2021iff})
uses the real part of the full effective potential, along with the tree-level kinetic terms of the nucleating fields, as the basis of the calculation of the critical bubble and bubble nucleation rate.
This approach is not self-consistent on several counts: (i) the DoFs of the nucleating fields are integrated over twice, (ii) an uncontrolled derivative expansion is implicitly relied upon and (iii) the imaginary part of the effective potential must be thrown away by hand.%
\footnote{
    This last point is often carried out without comment (for an exception see Ref.~\cite{Delaunay:2007wb}).
    For example, within the numerical package {\tt CosmoTransitions}~\cite{Wainwright:2011kj} the imaginary part of the potential is removed through the replacement $\log(m^2) \to \log(|m^2|)$ within the {\tt generic\_potential} module.
}
The EFT approach presented in this paper resolves all of these issues.

\approachtitle{Informal effective field theory approach}
Without the formal construction of an effective theory, the observation that some fluctuations should enter the effective action, and others should enter the statistical prefactor, has been utilised by several authors in concrete calculations of the bubble nucleation rate~\cite{Buchmuller:1992rs,Buchmuller:1993bq,Bodeker:1993kj,Gleiser:1993hf,Moore:1995jv,Ai:2018rnh,Ekstedt:2021kyx}.
Our approach formalises these ideas, and makes it unambiguous which fluctuations enter where (see Eq.~\eqref{eq:SigmaModes}).

\approachtitle{Functional renormalisation group approach}
This approach was pioneered in Ref.~\cite{Alford:1993br}, and was further explored in Refs.~\cite{Berges:1996ib,Berges:1996ja,Litim:1996nw,Strumia:1998nf,Strumia:1998qq,Strumia:1998vd,Munster:2000kk,Strumia:1999fq}, and more recently extended in Ref.~\cite{Croon:2021vtc}.
The general approach utilises functional renormalisation group methods
to compute the coarse-grained effective action, from which to solve for the critical bubble.
Thus, in spirit, it is very close to our approach.
In fact, the functional renormalisation group approach also depends upon a hierarchy of scales.
This is for two reasons.
First, at a fundamental level, the energy scale of fluctuations integrated out to define the coarse-grained effective action must be higher than the energy scale on which the non-convex effective potential flattens and becomes convex~\cite{Berges:1996ib,Berges:1996ja,Litim:1996nw} (this issue is resolved in Ref.~\cite{Croon:2021vtc}).
Second, at a practical level, because it is necessary to make a local approximation to the effective action, and in the absence of a hierarchy of scales this local approximation is not justified.
In addition, all the above studies which have utilised the functional renormalisation group approach to study thermal bubble nucleation have made use of high-temperature dimensional reduction to integrate out the non-zero Matsubara modes, with the functional renormalisation group used only when the UV and IR descriptions are assumed to contain the same field content.
In fact, regarding this last point, from an EFT perspective we can understand and overcome an apparent limitation of the functional renormalisation group approach.
In Ref.~\cite{Strumia:1998vd}, which studied the cubic anisotropy model (our Example 3), it was found that the saddle-point approximation broke down due to the large fluctuation determinants of the inducing $\chi$ field.
However, from an EFT perspective we have shown that the $\chi$ field should be integrated out into the nucleation scale effective action, and should not be included in the fluctuation determinants; see Eq.~\eqref{eq:SigmaModes}.
In other words, the $\chi$ mass threshold was not accounted for in Ref.~\cite{Strumia:1998vd}.
Doing so rescues the artificial breakdown of the saddle-point approximation, by including all exponentially large contributions into the nucleation action.

\approachtitle{Iterative nonlocal approach}
In this approach~\cite{Surig:1997ne,Garbrecht:2015yza} (see also Refs.~\cite{Ai:2018guc,Ai:2020sru}), one solves for the bubble (or bounce) from the equations of motion of the full nonlocal effective action, truncated at some loop order.
Starting from a bubble which solves some approximate local equations of motion, the nonlocal tadpole corrections are included iteratively.
If a hierarchy of scales exists, this approach is equivalent to ours, up to terms which are suppressed by powers of the ratio of scales.
This is because the hierarchy of scales ensures that there will exist an approximate local description for the full nonlocal effective action.
On the other hand, if there is no such hierarchy of scales, the iterative nonlocal approach is still, in principle, applicable.
Whether or not it is applicable in practice depends on the iterations converging~\cite{Baacke:2004xk,Baacke:2006kv}, which in turn depends on the accuracy of the initial estimate for the bubble. 
Note that these methods can be used in conjunction: Creating a nucleation scale EFT, within which one uses the iterative nonlocal method, already handles analytically and consistently large contributions for the higher scales, leaving only the nucleation scale and lower energy scales for numerical treatment.

\approachtitle{Lattice approach}
This approach was developed in Refs.~\cite{Moore:1998swa,Moore:2000jw,Moore:2001vf} (see also Ref.~\cite{Alford:1993ph} in lower dimensions), and entails a fairly direct lattice Monte-Carlo evaluation of the bubble nucleation rate.
As such it does not rely on the saddle-point approximation, nor on the applicability of perturbation theory within the nucleation scale effective theory.
So, in our Examples 2 and 3 it can be applied successfully to rather weak transitions.
However, the lattice approach does rely on an effective classical description of the nucleation scale, and hence on high-temperature dimensional reduction.
Further, where there is a hierarchy of scales, large lattices, perhaps prohibitively large, are required to resolve them.
As such, effective field theory methods play a crucial role in the lattice approach.

\vspace{0.2cm}

In summary, our approach resolves the inconsistencies of the naive and other approaches, and
agrees with other self-consistent approaches when there is a hierarchy of scales.
Only the iterative nonlocal approach and the lattice approach can in principle go beyond ours and apply where there is no hierarchy of scales.
However, the former is in practice limited by the convergence properties of the iterative scheme,
and the latter is limited by computational time, which restricts its use to the study of selected benchmark points.
Thus, the effective field theory approach to thermal bubble nucleation fills an important role:
It is the only self-consistent semiclassical method with a demonstrably local bounce equation.

Planned gravitational wave experiments, such as LISA, underline the importance of understanding cosmological first-order phase transitions reliably and quantitatively.
As we have argued in this article, effective field theory is an invaluable tool for self-consistently computing the thermal bubble nucleation rate.
However, the usefulness of effective field theory goes beyond this, as it also facilitates order-by-order gauge-invariant and renormalisation scale-independent calculations of equilibrium thermodynamic quantities, such as the free energy, critical temperature and latent heat~\cite{Farakos:1994kx,Braaten:1995cm,Kajantie:1995dw,Croon:2020cgk}.
Thus, effective field theory provides a general framework for reliably studying cosmological first-order phase transitions.

For the future, we leave two important extensions of the EFT approach to thermal bubble nucleation:
application to well-motivated models beyond the Standard Model
and the calculation of the dynamical prefactor.
While we have demonstrated the EFT approach to thermal bubble nucleation in only a few simple models, chosen to exhibit various characteristic behaviours, EFT provides a powerful tool that can be applied much more generally.
Models giving rise to a first-order electroweak phase transition are of particular interest, as these may be probed both at gravitational wave experiments~\cite{Audley:2017drz,Kawamura:2011zz,Harry:2006fi,Guo:2018npi},
and at the LHC or future colliders~\cite{Ramsey-Musolf:2019lsf}.
Analysis of gauge-Higgs symmetry breaking transitions can be carried out on the basis of Sec.~\ref{sec:scaleshifters}, while multi-step electroweak transitions may involve elements of all three examples, and more.
The latter extension, the calculation of the dynamical prefactor, requires directly tackling the nonequilibrium, real-time dynamics of an expanding bubble interacting with a thermal bath.
If this can be described by an effective Langevin equation for the infrared degrees of freedom, then the results of Refs.~\cite{Langer:1969bc,Berera:2019uyp} apply, yielding the dynamical prefactor in terms of the relevant dissipation coefficient.


\begin{acknowledgments}
We would like to thank
D.~Croon,
A.~Ekstedt,
H.~Gies,
J.~L\"{o}fgren,
P.~Millington,
L.~Niemi,
M.~J.~Ramsey-Musolf,
K.~Rummukainen,
P.~Schicho,
T.~V.~I.~Tenkanen,
and A.~Vuorinen
for enlightening discussions and valuable comments on the manuscript.
O.G.~(ORCID 0000-0002-7815-3379) was supported by the Research Funds of the University of Helsinki, and U.K.~Science and Technology Facilities Council (STFC) Consolidated grant ST/T000732/1.
J.H.~(ORCID 0000-0002-5350-7556) was supported by Academy of Finland grant no.~1322507, and the European Research Council grant no.~725369.
\end{acknowledgments}

\appendix


\section{Notation} \label{appendix:notation}

Our notation for momenta and loop integration follows Ref.~\cite{Braaten:1995cm,Braaten:1995jr}.
In particular, thermal four-momenta are denoted by uppercase letters, $P=(p_0,\mb{p})$, their components being the Matsubara frequencies, $p_0=\omega_n$, and the spatial momenta, $\mb{p}$.
Their norms squared are $P^2=p_0^2 + p^2$, where $p^2\equiv\mb{p}\cdot\mb{p}$. The loop integration measure in $d=3-2\epsilon$ dimensions is defined as,
\begin{equation}
\int_p \equiv \left(\frac{e^{\gE }  \blam^2 }{ 4 \pi }\right)^\epsilon \int \frac{d^{3-2\epsilon}p}{(2\pi)^{3-2\epsilon}}
\,,
\end{equation}
where $\gE$ denotes the Euler-Mascheroni constant.
Note that we have included powers of the \MSbar\ renormalisation scale, $\blam$, to make the measure up to mass dimension 3.
Sum-integrals over bosonic loop momenta are then defined as,
\begin{equation}
\sumint{P} \equiv T \sum_{p_0 = \omega_n^{\text{b}}} \int_p
\,,
\end{equation}
where the sum over $p_0$ implies a sum over integers $n$.
Sum-integrals over fermionic loop momenta are distinguished by curly braces around the loop momenta,
\begin{equation}
\sumint{\{P\}} \equiv T \sum_{p_0 = \omega_n^{\text{f}}} \int_p
\,.
\end{equation}
We also utilise the Feynman slash for contraction with the Euclidean gamma matrices, $\slashed{P}=P_\mu\gamma_\mu$.

Finally, we use the natural units common in particle physics, such that $c=\hbar=k_B=1$.


\section{Thin-wall regime} \label{appendix:thinwall}

The thin-wall limit of bubble nucleation arises as the temperature approaches the critical temperature (from the metastable phase).
The smallness of the deviation from the critical temperature results in a new hierarchy of scales, with the bubble radius, $R$, becoming much larger than the inverse mass of the nucleating field, $m_{\text{nucl}}$, which determines the thickness of the bubble walls.
The new hierarchy of scales $1/R \ll m_{\text{nucl}}$ allows for an extra step of the effective description: integrating out the scale $m_{\text{nucl}}$.
In fact this step is usually necessary when sufficiently deep in the thin-wall regime, which we will call the {\em strong thin-wall} regime.
Additionally, the estimate of the bubble volume made in Eq.~\eqref{eq:BubbleVolume} is no longer valid, with the consequence that additional exponential contributions to the nucleation rate arise.

The free energy of a thin-walled bubble can be given phenomenologically by~\cite{Gibbs1876}
\begin{equation}\label{eq:phenoThinWall}
    F_{\mathrm{bubble}} = \sigma A - p V \,,
\end{equation}
where $\sigma$ is the surface tension,
$p$ is the pressure difference across the bubble wall,
and $A$ and $V$ are the area and volume of the bubble wall respectively.
The pressure is equal to the difference in the free energy density $p = - \Delta f$, and in thin-wall regime this is approximately
\begin{equation} \label{eq:pressureLatentHeat}
p \approx \left(1-\frac{T}{T_c}\right)l \,,
\end{equation}
where $l=T_c d \Delta f/dT|_{T_c}$ is the latent heat.

Given spherical symmetry, we can solve for the radius of the critical bubble:
\begin{equation}\label{eq:ThinWallExtrem}
    \dv{R}F_{\mathrm{bubble}}=0 \Rightarrow R=\frac{2\sigma}{p}.
\end{equation}
In the approach to the critical temperature, $p$ becomes arbitrarily small.
The cost of the walls is proportional to the surface area of a bubble, but the effect of the interior free energy is proportional to the volume.
To compensate the cost of the walls, the critical bubble has to have a large interior consisting of the true new phase, and hence a large radius, so that the two effects are balanced.

Although the bubble radius grows in the thin-wall limit, the characteristic size of the wall width is still given by the mass of the nucleating field $d_{\mathrm{wall}}\sim m_{\text{nucl}}^{-1}$.
Thus, the thin-wall regime is 
\begin{equation}
    R = \frac{2\sigma}{p} \gg m_{\text{nucl}}^{-1} \,.
\end{equation}

In the thin-wall regime, the nucleation scale is no longer $m_{\text{nucl}}$, as was identified in Eq.~\eqref{eq:gradientSimMass}, but is an independent scale related to the bubble radius, $\Lambda_{\rmi{nucl}}\sim R^{-1}$.
Hence, the EFT for the scale $m_{\text{nucl}}$ still contains shorter scales than the nucleation scale.

To create the nucleation scale effective description, one has to integrate out the new intermediate scales at $R^{-1}\ll\Lambda_{\rmii{NI}}\lesssim m_{\text{nucl}}$. The wall thickness belongs to an intermediate scale ($d_{\mathrm{wall}}\sim m_{\text{nucl}}^{-1}$) and is thus absent in the effective description. This means that the nucleation scale contains only infinitely thin bubble walls as its dynamical DoFs.

The effective description for the bubble walls can be constructed based on general effective-theory principles.
First, one writes down the most general Lagrangian describing a two-dimensional closed membrane, $\mc{M}$, in three spatial dimensions which is invariant under reparameterisations of the membrane, and is a scalar under the three-dimensional Euclidean symmetry group.
Denoting by $\xi_a$, $a\in\{1,2\}$, the coordinates on the membrane,
the DoFs of the effective theory are the locations in space of the points on the membrane, $x^i(\xi_a)$, $i\in\{1,2,3\}$.
Then, as usual, one must match the long-wavelength predictions of the full theory to those of the effective theory.

The effective theory is constructed as a dual expansion in powers of couplings and in powers of the ratio of scales $1/(m_{\text{nucl}}R)$.
The operators of the membrane theory are ordered according to increasing mass dimension, and the leading few coefficients are matched to reproduce the infrared predictions of the UV theory,
\begin{align} \label{eq:actionThinWall}
\F [\mc{M}] = \int_{\mc{M}} \sqrt{\gamma}\ \dd^{2}\xi \left[- \frac{1}{3}p x^i n_i +  \sigma  + c \mc{K} + \dots \right] \,.
\end{align}
Here $\gamma$ is the determinant of the induced metric on the membrane $\gamma_{ab}=\partial_a x^i\partial_b x_i$,
$n_i$ is the unit normal pointing out of the bubble,
$\mc{K}$ is the extrinsic curvature of the membrane,
and $p$, $\sigma$ and $c$ are constant coefficients.
The first two terms in Eq.~\eqref{eq:actionThinWall} reproduce Eq.~\eqref{eq:phenoThinWall}, though the volume term has been be rewritten as an integral over $\mc{M}$ using Gauss's law~\cite{Guenther:1979td,Garriga:1991ts}, so that Eq.~\eqref{eq:actionThinWall} appears as a Lagrangian theory.

If this effective description is to prove fruitful, the additional terms suggested by the ellipsis in Eq.~\eqref{eq:actionThinWall} should be of higher mass dimension, containing additional derivatives of the membrane position or normal vector.
From standard effective theory arguments, we would expect that the magnitude of the constant coefficients, matched to the UV theory, is determined by a UV mass scale ($m_{\text{nucl}}$ or larger) to the appropriate power, excepting $p$ which is anomalously small in the thin-wall regime.
On the other hand, the membrane operators must be of order $R$ to the appropriate power, as this is the only scale present in the membrane theory.

In the thin-wall regime, the partition function of the full theory reduces to that of the membrane theory, up to an overall constant,
\begin{equation}
    Z(T) \propto \int^{(\Lambda)} \mc{D M}\exp\big\{ -\beta \F [\mc{M}]\big\} \,,
\end{equation}
where $\Lambda$ is now the matching scale between the scales $1/R$ and $m_{\text{nucl}}$.
For the leading non-trivial theory, i.e.\ only $p$ and $\sigma$ nonzero, the one-loop statistical prefactor for this theory has been calculated in Refs.~\cite{Guenther:1979td,Garriga:1993fh}.%
\footnote{
    Note that the extrinsic curvature contribution to $\beta \F $ is linear in $R$ (see also Refs.~\cite{Ignatius:1993zq,Widyan:2012ad}), and hence is larger than the prefactorial corrections calculated in Refs.~\cite{Guenther:1979td,Garriga:1993fh} when deep in the thin-wall regime.
}

It is not always an imperative to construct the effective description for the bubble walls, if the system is not in the strong thin-wall regime.
The new intermediate scale can be treated perturbatively if it has a small effect on the free energy describing the critical bubble.
Crucially, the dominant contribution to the surface tension, $\sigma$, must come from higher scales still, as the wall width itself is of order $m_{\text{nucl}}^{-1}$.
If, in addition, the new intermediate scale gives only a small correction to the pressure
\begin{equation}\label{eq:pressureCondition}
    \Delta p_{\rmii{NI}}\ll p \,,
\end{equation}
then it is possible to calculate the bubble nucleation rate either within the thin-wall effective theory, or within the EFT at the scale $m_{\text{nucl}}$.
In this case the thin-wall effective theory may be useful, but is not necessary.
When using the EFT at the scale $m_{\text{nucl}}$, one can account for the effect of the new intermediate scales on the shape of the critical bubble by computing their tadpole contributions, as discussed around Eqs.~\eqref{eq:shapechangeonelooporigins}--\eqref{eq:shapechangecontrib}.
See Ref.~\cite{Bezuglov:2018qpq} for an example of this in the context of vacuum decay.

On the other hand, for temperatures sufficiently close to the critical temperature (where $p=0$), Eq.~\eqref{eq:pressureCondition} typically breaks down, as even a perturbatively suppressed correction can be of LO for the pressure difference.
This, in turn, will have a LO effect on the critical bubble (\textit{cf.} Eq.~\eqref{eq:ThinWallExtrem}), thus compromising the perturbative expansion.
This is the strong thin-wall regime.

In the strong thin-wall regime, the construction of the thin-wall effective theory is still possible, as a dual expansion in the couplings and in the deviation from the critical temperature.
Deep in the thin-wall regime, corrections due to the finiteness of the deviation from the critical temperature are subdominant, and one can work to leading nontrivial order in $|1-T/T_c|$, while working to higher order in the couplings.
That is, one calculates the surface tension at $T_c$, and the pressure according to Eq.~\eqref{eq:pressureLatentHeat}.
Corrections to the pressure from the new intermediate scales can be calculated about a homogeneous background, but corrections to the surface tension from new intermediate scale fluctuations must be calculated in the bubble wall background (see for example Refs.~\cite{Dashen:1974cj,Garbrecht:2015oea,Bezuglov:2018qpq}).
In order to calculate the exponent of the nucleation rate to $\order{1}$ accuracy, one must calculate the pressure, surface tension and extrinsic curvature coefficient $c$ to within uncertainties of order $T/R^3$, $T/R^2$ and $T/R$ respectively.

The strong thin-wall regime does not arise in the special case whereby an exact symmetry determines $T_c$, such as for the 3d EFT of Sec.~\ref{sec:one_scalar} where a $\mathbb{Z}_2$-symmetry relates the two phases at $T_c$.
Due to this symmetry, the condition in Eq.~\eqref{eq:pressureCondition} is not broken, and hence the strong thin-wall regime does not arise.
This special case is, in fact, the model for which the thin-wall regime has been most widely studied~\cite{Coleman:1977py,Ignatius:1993zq,Widyan:2012ad}.

As a closing aside in the context of the thin-wall approximation, we would like to highlight the work of Refs.~\cite{langer1973hydrodynamic,kawasaki1975growth,Turski1980,Csernai:1992tj,Venugopalan:1993vk,Carrington:1993ng}, in which an effective hydrodynamic description was assumed to apply at the nucleation scale, containing thin-wall bubbles.
Within this effective description, the nucleation rate has been calculated explicitly, including the statistical and dynamical prefactors.
The latter was found to depend on the bulk and shear viscosities of the fluid.
However, a first-principles derivation of this effective hydrodynamic description, including of its region of validity for bubble nucleation, is still lacking.
This was seen in Ref.~\cite{langer1973hydrodynamic} as {\em a serious gap in our logical development}.
Filling this gap in full detail would be a very interesting extension of our work.


\section{Not a high-temperature phase transition} \label{appendix:lowerTemp}

In the EFT approach to thermal bubble nucleation, the thermal scale is integrated out in dimensional reduction, removing the Euclidean time dimension from the description.
This step is necessary for the link to classical nucleation theory, as discussed in Sec.~\ref{sec:background}.
However, it is only possible if $\Lambda_{\mathrm{nucl}}\ll\Lambda_{\mathrm{therm}}$, so that the bubbles are much larger than the extent of the Euclidean dimension.
Clearly this is not the only possible case, as there may be transitions with $\Lambda_{\mathrm{nucl}}\sim\Lambda_{\mathrm{therm}}$ or $\Lambda_{\mathrm{nucl}}\gg\Lambda_{\mathrm{therm}}$, which we will refer to as {\em intermediate} and {\em low} temperatures respectively.

At zero temperature, vacuum decay through bubble nucleation is well understood.
In this case the real-time process of quantum tunnelling can be related to the imaginary part of the vacuum energy, $\Gamma_{\rm vac}=-2\Im E_{\rm vac}$~\cite{Schwinger:1951nm} (see also Ref.~\cite{Andreassen:2016cff} for an approach avoiding analytic continuation).
This in turn can be calculated in a purely Euclidean setting, which is simpler still than classical nucleation theory, in which the dynamical prefactor (see Eq.~\eqref{eq:divisionDynamicalStatistical}) requires a real-time calculation.

It is worth noting that the EFT approach to bubble nucleation is also useful at zero temperature, in particular for radiatively-induced vacuum transitions (see for example Refs.~\cite{Weinberg:1992ds,Andreassen:2016cvx}).
By first integrating out the high-momentum modes, to arrive at an EFT for the nucleating DoFs, the problems of double-counting DoFs, stray imaginary parts and an uncontrolled derivative expansion can be avoided, and the vacuum decay rate calculated in a consistent power expansion.

Without the hierarchy $\Lambda_{\mathrm{nucl}}\ll\Lambda_{\mathrm{therm}}$, an effective classical description is not in general possible, in which case the real-time dynamics of the quantum theory must be tackled head on.
While at zero temperature the rate of vacuum decay can be related to the imaginary part of the vacuum energy, this is not the case at nonzero temperature and hence one must return to the Schwinger-Keldysh formalism~\cite{Chou:1984es} to formulate the problem of bubble nucleation.

Nevertheless, Linde's conjectured analogy $\Gamma \sim -2\Im F$~\cite{Linde:1980tt,Linde:1981zj} may still capture the qualitative temperature dependence of the exponent of the nucleation rate, indeed it agrees with the quantum mechanical escape rate at low temperatures~\cite{Affleck:1980ac}.
In this approach, the relevant saddle-point solution determining the nucleation process changes from being $\mathrm{O}(4)$ to $\mathrm{O}(3)$ to $\mathrm{O}(3)\times \mathrm{O}(2)$ symmetric as the temperature increases from zero through the low, intermediate and high temperature regimes.%
\footnote{
    Strictly speaking the $\mathrm{O}(4)$-symmetric solution is only applicable at $T=0$, however for theories without massless excitations at $T\neq 0$ the action of the $\mathrm{O}(4)$-symmetric solution is equal to that of the $\mathrm{O}(3)$-symmetric one up to exponentially small corrections, until the temperature raises to approximately $1/(2R)$, where $R$ is the radius of the bubble.
}
Studies within this approach have found that the intermediate $\mathrm{O}(3)$ case may only occur in a rather fine-tuned range of temperatures~\cite{Garriga:1994ut,Ferrera:1995gs,Widyan:1998wa,Salvio:2016mvj}; see also Ref.~\cite{Abed:2020lcf} for a recent study.
In fact, in the thin-wall regime this intermediate case does not occur at all~\cite{Garriga:1994ut}, and the nucleation rate jumps from vacuum tunnelling to an over-barrier process as the temperature is increased.


\section{Numerical methods} \label{appendix:numerics}

A range of numerical techniques have been developed for concrete calculations of the factors entering Eq.~\eqref{eq:SaddlepointAppr}.
Following Ref.~\cite{Coleman:1977py}, the shooting method has been widely and successfully used to solve for the critical bubble.
However, for theories with multiple scalar fields, the convergence of this method is not guaranteed, so a variety of methods have been developed to overcome or ameliorate this~\cite{Cline:1999wi,Konstandin:2006nd,Wainwright:2011kj,Akula:2016gpl,Masoumi:2016wot,Espinosa:2018hue,Espinosa:2018szu,Piscopo:2019txs,Sato:2019axv,Hirvonen:2020jud,Bardsley:2021lmq}.

The functional determinant, which makes up the statistical prefactor, is radially separable and can be decomposed using spherical harmonics.
The computation of the radial part can then be simplified by making use of a classic result of Gel'fand and Yaglom~\cite{Gelfand:1959nq};
see for example Refs.\cite{Baacke:1993ne,Baacke:2003uw,Dunne:2005rt,Dunne:2006ct,Dunne:2007rt,Ekstedt:2021kyx}.
In Ref.~\cite{Ekstedt:2021kyx} functional determinants have been calculated for several theories within a 3d EFT framework.
The numerical results of this work can in fact be used directly to obtain the statistical part of the nucleation rate for our Example 1.

\bibliography{refs}
\end{document}